\documentclass[12pt,a4paper]{article}
\pdfoutput=1
\usepackage{jheppub}
 \usepackage{amsmath}
 \usepackage{amsfonts}
 \usepackage{mathtools}
 \usepackage{amssymb}
 \usepackage{caption}
 \usepackage{subcaption}
 \usepackage{slashed}

 \title{Constraints on parity violating conformal field theories in $d=3$}
  \author{Subham Dutta Chowdhury ${}^a$, Justin R. David ${}^a$, Shiroman Prakash ${}^b$}
\affiliation{ ${}^a$ Centre for High Energy Physics, Indian Institute of Science,\\
C. V. Raman Avenue, Bangalore 560012, India.\\
${}^b$Department of Physics and Computer Science, \\
Dayalbagh Educational Institute, Dayalbagh, \\
Agra 282005, India.}
\emailAdd{subham, justin@cts.iisc.ernet.in, shiroman@gmail.com}
\abstract{ 
We derive constraints on three-point functions involving the stress
tensor, $T$, and a conserved $U(1)$ current, $j$, in 2+1 dimensional
conformal field theories that violate parity, using conformal collider
bounds introduced by Hofman and Maldacena. Conformal invariance allows
parity-odd tensor-structures for the $\langle T T T \rangle$ and $
\langle j j T \rangle$ correlation functions which are unique to three
space-time dimensions. Let the parameters which determine the $\langle
T T T \rangle$ correlation function be $t_4$ and $\alpha_T$ , where
$\alpha_T$ is the parity-violating contribution. Similarly let the
parameters which determine $ \langle j j T \rangle$ correlation
function be $a_2$, and $\alpha_J$ , where $\alpha_J$ is the
parity-violating contribution. We show that the parameters $(t_4,
\alpha_T)$
and $(a_2, \alpha_J)$ are bounded to lie inside a disc at the origin
of the $t_4$ - $\alpha_T$ plane and the $a_2$ - $\alpha_J$ plane
respectively. We then show that large $N$ Chern-Simons theories
coupled to a fundamental fermion/boson lie on the circle which bounds
these discs. The `t Hooft coupling determines the location of these
theories on the boundary circles.
}

\begin{document}
\maketitle
\section{Introduction}

Conformal field theories in $d=3$ are of interest both in the context of 
holography as well as in condensed matter physics. In the context of holography there are several well studied 
CFTs in $d=3$ which are known to admit a holographic dual. 
Just to name a few, the  M2-brane theory \cite{Maldacena:1997re}, the free and critical 
$O(N)$ model \cite{Klebanov:2002ja,Leigh:2003gk,Sezgin:2003pt}, 
the ABJ(M) theories \cite{Aharony:2008ug,Aharony:2008gk} and more recently the 
Chern-Simons theories with fundamental matter \cite{Giombi:2011kc,Aharony:2011jz}. 
The critical $O(N)$ model as well as the Chern-Simons theories with fundamental matter
are proposed to be dual to higher spin theories in $AdS_4$. 
Several super conformal field theories have been 
constructed in  \cite{Gaiotto:2008cg} which exhibit a rich structure of dualities \cite{Giveon:2008zn}. Holographic theories have been used
study strongly coupled phenomenon relevant to condensed matter 
physics \cite{Hartnoll:2007ih}. Conformal field theories in $d=3$ exhibit a  rich variety of physical phenomena, 
and a detailed study of them can provide an understanding of quantum gravity in four dimensions.

A particularly important feature of conformal field theories in three dimensions is that physically 
relevant examples need not preserve parity, thanks to the possibility of a 
Chern-Simons term \cite{Deser:1981wh, Deser:1982vy, Frohlich:1989gr}. The study of Chern-Simons 
theories with matter in the past several years has been very fruitful, and led to the discovery of 
several new non-supersymmetric dualities: the higher spin/vector model 
duality, see \cite{Giombi:2016ejx} for a review and the Bose-Fermi duality \cite{Aharony:2012nh}, which 
has been well tested at large $N$ 
\cite{GurAri:2012is, Aharony:2012ns, Jain:2013py, Bedhotiya:2015uga, Gur-Ari:2015pca, Giombi:2016zwa}, and 
is also believed to hold at finite $N$ \cite{Aharony:2015mjs}. Quantum Hall 
fluids \cite{Frohlich:1990xz, Frohlich:1991wb, Zee} are a particularly important physical example
where such theories are relevant.

Consider a conformal field theory in $d=3$ with a  $U(1)$ conserved current $j$. 
Associated with the conformal field theory, is its stress tensor $T$ and let the theory be parity violating. 
In such a theory,   conformal invariance constrains 
the  three point functions $\langle jjT \rangle$ and $\langle TTT \rangle$
to be of the form  \cite{Giombi:2011rz,Maldacena:2011jn},
\begin{eqnarray}\label{threepoint function}
\langle jjT \rangle &=& n^j_s\langle jjT \rangle_{\textrm{free boson}} 
+n^j_{f}\langle jjT \rangle_{\textrm{free fermion}} + p_j \langle jjT \rangle_{\textrm{parity odd}},\\ 
\langle TTT \rangle &=& n^T_s\langle TTT \rangle_{\textrm{free boson}} 
+n^T_{f}\langle TTT \rangle_{\textrm{free fermion}} + p_T \langle TTT \rangle_{\textrm{parity odd}},\nonumber  
\end{eqnarray}
where $\langle .. \rangle_{\textrm{free boson}}$, 
$\langle .. \rangle_{\textrm{free fermion}}$ denote the correlator  a real free  boson and a real 
free fermion respectively.  
The parity even tensor structures are written down in \citep{Osborn:1993cr}, 
while the parity odd structure 
was first discovered  in \citep{Giombi:2011rz}.
The  numerical coefficients  $n_s^{j, T}, n_f^{j, T}$ are theory dependent, and once
the normalization of the parity odd term is fixed, the parity violating 
coefficient $p_{j, T}$ can be determined from a given theory \cite{Giombi:2011kc,Aharony:2011jz}
\footnote{Similar calculations of three point functions of 
conserved currents in supersymmetric Chern-Simons theories with matter 
has been done in \cite{Buchbinder:2015qsa,Buchbinder:2015wia,Kuzenko:2016cmf}. 
Such theories do not have the parity violating contribution. }.

For parity even theories and for theories in $d=4$ and higher dimensions the 
conformal collider bounds, found by \cite{Hofman:2008ar}, impose constraints on 
the parity even coefficients that occur in such correlators 
\cite{deBoer:2009pn,Buchel:2009sk,Camanho:2009vw}. 
However the role  of the parity odd coefficients $p_{j, T}$  
in conformal collider bounds in $d=3$ has not been investigated in detail. 
In the  appendix  C of \cite{Maldacena:2011jn}, using  
general symmetry arguments, the contribution of a 
 parity odd term to the energy flux in the conformal collider  for excitations created by the
 stress tensor 
 was written down. However the precise relation with the 
coefficient $p_T$ was not provided.

In this paper we start from the correlators given in 
\ref{threepoint function} and obtain the 
energy flux in the conformal collider as a function of the
parameters in the $3$ point function 
and impose the positive energy flux condition of Hofman and Maldacena.  On normalising the 
$2$-point function of the stress tensor and the  $U(1)$ current, 
the energy flux for excitations created by the stress tensor or the 
$U(1)$ current is determined by $2$ parameters. 
We show that the positive energy flux condition constrains these parameters to 
the region of a disc at the origin. 
We also show that large $N$ Chern-Simons theories lie on the bounding circle of
this disc. Their position on the circle is determined by the 't Hooft coupling.

\paragraph{  } To state our results, let us briefly describe the set up of \cite{Hofman:2008ar}.  
A gedanken collider physics experiment was carried out, where one 
studies the effect of localized perturbations at the origin. The integrated 
energy flux per unit angle, over the states created by such perturbations, was measured 
at a large sphere of radius $r$. \begin{eqnarray}
\langle {E}_{\hat n} \rangle &=& \frac{\langle 0|\mathcal{O}^\dagger{E}_{\hat n}
\mathcal{O}|0\rangle}{\langle 0|\mathcal{O}^\dagger \mathcal{O}|0\rangle},\nonumber\\                    
{E}_{\hat n}  &=& \lim_{r \rightarrow \infty} r^2 \int_{-\infty}^{\infty} dt n^iT^{t}_i(t, r\hat{n}),\nonumber\\
\mathcal{O} &\sim & \frac{\epsilon^{ij} T_{ij}}{\sqrt{\langle \epsilon^{ij}T_{ij}|T_{ij}\epsilon^{ij}\rangle}}, \qquad \frac{\epsilon^{i} j_{i}}{\sqrt{\langle \epsilon^{j}j_{j}|j_{i}\epsilon^{i}\rangle}},
\end{eqnarray}                       
where, $\hat n$ is a unit vector in $R^3$, which specifies the direction of the 
calorimeter and $\mathcal{O}$ is the operator creating the localised perturbation.  Under a suitable 
transformation of coordinates it can be shown that positivity of energy flux measured 
in such a way is equivalent to demanding that the averaged null energy taken over the states be positive,
The fact that averaged null energy is positive in any unitary interacting conformal 
field theory was shown in \citep{Faulkner:2016mzt,Hofman:2016awc,Hartman:2016lgu}. Thus, by 
demanding the positivity of energy flux, one obtains various constraints on the parameters 
of the three point function of the CFT, depending on the operators used to create the states $\mathcal{O}$. Similar constraints for correlators involving higher 
spins were obtained using unitarity in \cite{Komargodski:2016gci}. 
 There has been a systematic study of such constraints both from the context of holography and CFT.  Recently such constraints were used to place bounds on 
 spectral sum rules of CFTs in arbitrary dimensions \cite{Chowdhury:2016hjy}. 
 
\paragraph{  } In this paper we  perform this analysis for 
 general CFTs in $d=3$ including the parity odd terms in 
(\ref{threepoint function}). 

For states created by insertion 
of currents in two orthogonal directions say $x, y$ and the 
calorimeter placed in the $y$ direction, the energy  observed in the conformal collider 
 takes the form of a matrix which is given by 
\begin{eqnarray}
\hat{E}(j)&=&  \begin{pmatrix}
\frac{E}{4\pi }(1-\frac{a_2}{2}) &&\frac{ \alpha_j E}{8\pi}\\
 \\
\frac{ \alpha_j E}{8\pi}&& \frac{E}{4\pi }(1+\frac{a_2}{2})\\
 \end{pmatrix},
\end{eqnarray}
where $a_2$ is the dimensionless parameter  introduced by 
Hofman and Maldacena \cite{Hofman:2008ar} and 
$\alpha_j$ is the contribution of the parity odd 
part of the three point function to the energy functional. 
They are related to the three parameters of the three point functions as follows
\begin{eqnarray}\label{reljmh}
a_2 = -\frac{2 (n^j_f-n^j_s)}{(n^j_f+n^j_s)}, \qquad   \alpha_j = \frac{4 \pi ^4 p_j}{(n^j_f+n^j_s)}.
\end{eqnarray}

The diagonal elements of the energy  matrix are  due to the 
parity even part of the three point correlators while, parity odd contributions 
to the three point functions is responsible for the off diagonal elements.  
A similar matrix 
is obtained when one considers the localized perturbations  created by stress tensor insertions.
This is given by 
  \begin{eqnarray}
 \hat{E}(T) = \begin{pmatrix}
 \frac{E}{4\pi} (1-\frac{t_4}{4}) &&\frac{ \alpha_T E}{16\pi} \\
 \\
\frac{ \alpha_T E}{16\pi} && \frac{E}{4\pi}(1+\frac{t_4}{4})\\
 \end{pmatrix},
\end{eqnarray}   
where $t_4$ is the dimensionless parameter first introduced by 
\cite{Hofman:2008ar}  and $\alpha_T$ is  
the parity odd contribution.  They are 
related to the  coefficients in (\ref{threepoint function})  by 
\begin{eqnarray}\label{reltmh}
t_4 = -\frac{4 (n^T_f-n^T_s)}{n^T_f+n^T_s}, \qquad 
\alpha_T =  \frac{8 \pi ^4 p_T}{3 (n^T_f+n^T_s)}.
\end{eqnarray}

 Positivity of energy  requires that the eigenvalues of these 
 symmetric matrices be positive. Since the trace of the diagonal elements is 
 always positive, this condition boils down 
 to the fact that the determinant must be positive. This leads us to the following constraints
\begin{eqnarray}\label{disc}
a_2^2 + \alpha_j^2 \leq 4 \nonumber, \qquad 
t_4^2 + \alpha_T^2 \leq 16.
\end{eqnarray}
Thus the $2$ parameters  determining each of the $3$ point functions 
are constrained to lie on a disc at the origin. 
 
\begin{figure}[h]
\center
\includegraphics[scale=0.7]{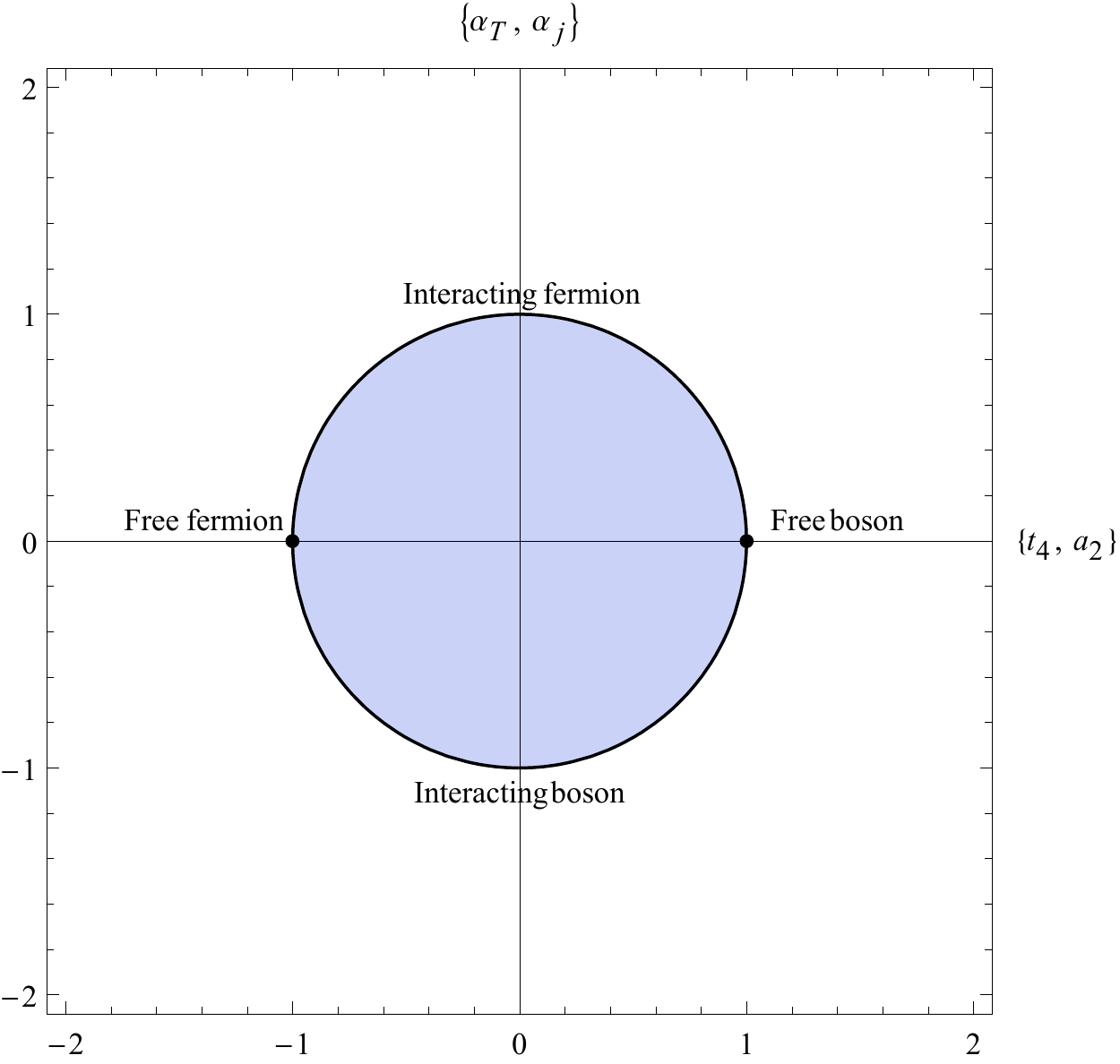}
\caption{The space of conformal field theories in $d=3$ obeying 
the conformal collider bounds is shaded. The $y$-axis denotes the 
parity odd coefficient of either the 3-pt functions $\langle TTT\rangle$ or 
$\langle jjT\rangle$, the $x$-axis denotes the parity 
even coefficient. Large $N$ Chern-Simons theories 
lie on the boundary of the disc. The position
on the disc corresponds to  the 't Hooft coupling of these 
theories. If we choose the convention that $\lambda$ is positive, the top-half circle corresponds to fundamental fermions, and the bottom half corresponds to fundamental bosons.}\label{bounds}
\end{figure}

\paragraph{} 
As discussed, 
large $N$ Chern Simons theories at level $\kappa$ coupled to fundamental matter 
are conformal field theories which are known to  be parity violating. 
Let us consider the case of fundamental fermions.
Using softly broken higher spin symmetry, it is known that the  theory dependent 
parameters of the three point functions in (\ref{threepoint function}) are given by 
\cite{Maldacena:2012sf} 
\begin{eqnarray}\label{cscoeffTTTi}
n^T_s(f) = n^j_s(f)  = 2N\frac{\sin \theta}{ \theta} \sin^2 \frac{\theta}{2}, &\qquad &
 n^T_f(f) = n^j_s(f) = 2N\frac{\sin \theta}{ \theta} \cos^2 \frac{\theta}{2},  \nonumber \\
 p_T(f)= \alpha N \frac{\sin^2 \theta}{\theta} , &\qquad &
p_j(f) = \alpha' N \frac{\sin^2 \theta}{\theta},
\end{eqnarray}
where $\theta$ is the 't Hooft coupling given by 
\begin{equation}
\theta = \pi \frac{N_f}{\kappa}.
\end{equation}
Here the $(f)$ in the brackets refer to the fact that we are dealing with the theory of 
fermions in the fundamental representation.  The dependence of the parameters
on the t 'Hooft coupling can also be found by summing over the planar diagrams
as done in \cite{Aharony:2012nh, GurAri:2012is}. 
Once the normalization of the parity odd tensor structures 
$\langle jjT\rangle_{\textrm{parity odd}}, \langle TTT\rangle_{\textrm{parity odd}}$
is agreed up on, the numerical constants $\alpha, \alpha'$ can  be determined by 
a perturbative one loop calculation.  
We fix the normalization of the parity odd tensor structures  as given in \cite{Giombi:2011kc}
\footnote{These
are written down in equations (\ref{jjt}) and (\ref{ttt}). 
Of course, one could instead choose to define the normalisation for
the parity odd forms of three-point functions by the convention that
$\alpha=\alpha'=1$. This would be the (possibly more natural) choice
of normalisation of parity-odd forms in which the results of
\cite{Maldacena:2012sf} are implicitly stated}. 
We then carefully redo 
the perturbative one loop analysis of \cite{Giombi:2011kc}  in section \ref{podd3pt} and 
find
\begin{equation}
\alpha =  \frac{3}{\pi^4}, \qquad \alpha' =  \frac{1}{\pi^4}.
\end{equation}
Now substituting these values in (\ref{reljmh}) and (\ref{reltmh}) we obtain 
\begin{eqnarray}
a_2 = -2 \cos\theta, \qquad \alpha_j = 2\sin\theta, \\ \nonumber
t_4 = -4 \cos\theta, \qquad \alpha_T = 4 \sin\theta.
\end{eqnarray}
Thus Chern-Simons theory with fundamental matter lie on the circle bounding the disc (\ref{disc}). 
Their location on the bounding circle is parametrized by the 
't Hooft coupling $\theta = \frac{\pi N}{\kappa}$. Here $\theta = 0$ is the theory with free 
fermions, while $\theta = \pi$ is the theory with critical bosons. 
The range $0<\theta <\pi$ can be thought of a theory with interacting fermions
\footnote{Note that in this convention, $\theta$ is measured from the negative $[t_4, a_2]$ axis. 
However if we flip the sign of $t_4, a_2$,  then $\theta$ would have the conventional defintion.}
in the fundamental representation with positive $\kappa$ (or interacting bosons in the fundamental representation with negative $\kappa$). The range $ \pi<\theta <2\pi$ corresponds to 
the theory of interacting bosons with positive $\kappa$ (or interacting fermions in the fundamental representation with negative $\kappa$). This is because using the 
bosonization map of \cite{Aharony:2012nh, GurAri:2012is}, 
the 't Hooft coupling of the interacting bosonic theory is 
related to the fermionic one by $\theta_b = \pi + \theta_f$. 
We summarise the space of conformal field theories in 
3 dimensions in figure \ref{bounds}

\paragraph{} The organization of the paper is as follows. In section \ref{setup}, we 
set up the kinematics for the energy matrix observed at the collider. 
In  section \ref{jjtcalculation} and \ref{tttcalculation} we evaluate the energy matrix for excitations
created by current and stress tensor respectively. We  show that 
demanding the positivity of the eigen values of the energy matrix, 
the parameters $a_2, \alpha_j$ and $t_4, \alpha_T$ are constrained to lie 
on a disc at the origin in their respective 2-planes. 
In section \ref{largeNchernsimons} we apply these results to 
large $N$ Chern-Simons theories coupled to fundamental matter and demonstrate 
that these theories lies on the circles bounding the discs. 
The position on the circle is determined by the 't Hooft coupling of the theory. 
Appendix \ref{details} contains the detail evaluation of one of the terms 
contributing to the energy matrix corresponding to stress tensor excitation
as an example. Appendix \ref{oneloopcalculation} contains the details of the 
perturbative evaluation of the parity odd 
 three point functions of interest in large $N$ Chern-Simons theories coupled
 to fundamental fermions. Finally appendix \ref{A} contains  the results of some 
 integrals which arise in  evaluation of the energy matrix.

\section{Energy matrix and positivity of energy }\label{setup}

In this section we set up the thought experiment in a bit more detail. Following \citep{Hofman:2008ar}, we consider localised perturbations of  the CFT in minkowski space.
\begin{eqnarray}
ds^2= -dt^2+dx^2+dy^2.
\end{eqnarray}
These perturbations evolve in time and spread out. In order to measure the energy flux, we consider concentric circles (concentric spheres for higher dimensions) at which the detector is placed. The energy  measured in a direction $\hat n$ is then defined as,

\begin{eqnarray}\label{energyfunctionaldef}
E_{\hat n} &=& \lim_{r \rightarrow \infty} r \int_{-\infty}^{\infty} dt n^iT^{t}_i(t, r \hat n),
\end{eqnarray}
where $r$ is the radius of the circle on which the detector is placed and $\hat n$ is a unit vector which determines the point on the circle where the detector is placed.  Alternatively one can place the detector at the future null infinity from the very beginning  and integrate over the null time. These two definitions are  equivalent \citep{Zhiboedov:2013opa}. For our convenience, we place the detector along the $y$ direction i.e $\hat n=(0,1)$.  In order to calculate the contributions to the energy  (eqn \ref{energyfunctionaldef}) at future null infinity, we introduce the light cone coordinates
$x^{\pm}=t\pm y$. Let $x$ be the other spatial co-ordinate. 
In the limit $r \rightarrow \infty$, eqn \ref{energyfunctionaldef} becomes,
\begin{eqnarray}\label{mathcale}
E &=& \lim_{x^{+} \rightarrow \infty} (\frac{x^+ - x^-}{2}) \int_{-\infty}^{\infty} \frac{dx^-}{2}(-T_{+y}-T_{-y})n^y, \nonumber\\
&=&   \lim_{x^{+} \rightarrow \infty} (\frac{x^+ - x^-}{2}) \int_{-\infty}^{\infty} \frac{dx^-}{2}(-T_{++}+T_{--}),
\end{eqnarray}
where we have dropped the subscript $\hat{n}$. 
To see that the contributions form $T_{++}$ vanish at future null  infinity $x^+\rightarrow 0$, 
we follow \cite{Hofman:2008ar} and perform the following  co-ordinate transformation
\begin{eqnarray}\label{cotrans}
y^+=-\frac{1}{x^+}, \qquad y^-= x^- - \frac{x^2}{x^+}, \qquad y^1 &=& \frac{x}{x^+}.
\end{eqnarray} 
The stress tensor changes as
\begin{eqnarray}
T^x_{\mu \nu} &=& \frac{\partial y^c}{ \partial x^\mu} \frac{\partial y^d}{ \partial x^\nu} T^y_{cd}, 
\end{eqnarray}
and using the transformation in (\ref{cotrans})  we obtain
\begin{eqnarray}
\lim_{x^+\rightarrow 0} T^x_{++} = (y^1)^4 T^y_{--}, \qquad T^x_{--} = T^y_{--}.
\end{eqnarray} 
Here superscripts refer to the stress tensor in the respective co-ordinate system. 
Thus at future null infinity we see that its only the component $T_{--}^y$ which determines
all the components of the energy in the $x$ co-ordinate system. 
Further more the component $T^x_{++} $ is suppressed. 
Therefore the energy detected at the calorimeter is given by 
\begin{eqnarray}
E &=&  \lim_{x^{+} \rightarrow \infty} (\frac{x^+ - x^-}{2}) \int_{-\infty}^{\infty} \frac{dx^-}{2}T_{--}. 
\end{eqnarray}
We are interested in the expectation value of the 
energy operator  on states created by stress tensor and current insertions.  
The normalized states are defined as 
\begin{equation}\label{state}
{\cal O} _E  |0\rangle = \frac{\int dt  dx dy e^{i E t} {\cal O}( t, x, y) |0\rangle}{
\sqrt {\langle  {\cal O}_E|  {\cal O}_E| \rangle} },
\end{equation}
where ${\cal O}$ are operators  constructed from the current or stress tensor
with definite polarizations. 
They are given by 
\begin{equation} \label{defotj}
{\cal O} (\epsilon;T) = \epsilon_{ij} T^{ij}, \qquad \hbox{or} \qquad  {\cal O}(\epsilon, j)  = \epsilon_i j^i. 
\end{equation}
The norm in (\ref{state}) is defined by 
\begin{equation}\label{defsotj}
\langle {\cal O}_E |  {\cal O}_E  \rangle = 
\int d^3 xe^{i E t} \langle {\cal O} (t, x, y) {\cal O}(0) \rangle,
\end{equation}
where we have used translation invariance to factor out one integral and place one
of the operators at the origin. 
Thus the norms are obtained by evaluating the two point function of the operator 
and integrating the space time point corresponding to one of the operator.

Let us first look at states created by the stress tensor excitations.  
Since the stress tensor is traceless, the allowed polarisations for stress tensors 
satisfy
\begin{eqnarray}
\epsilon^\mu_\mu = 0.
\end{eqnarray}
Therefore we choose $2$ independent polarisations given by 
\begin{eqnarray}
\epsilon^{xy}=\epsilon^{yx}=1, \nonumber\\
\epsilon^{\prime xx}=-\epsilon^{\prime yy}=1.
\end{eqnarray}
Let us label the states created by these polarizations as
\begin{equation}
{\cal O}_E(\epsilon; T) |0\rangle, 
  \qquad {\cal O}_E(\epsilon'; T) |0\rangle,
\end{equation}
where we use the definition of the state given in (\ref{defotj}) and (\ref{defsotj}) with 
the polarizations $\epsilon$ and $\epsilon'$. 
We can then define the energy matrix between these states as
\begin{eqnarray}\label{enmat}
\hat E(T) = 
\left( \begin{array}{cc}
\langle 0| {\cal O}_E^\dagger (\epsilon; T) \mathcal{E} {\cal O}_E (\epsilon; T) |0\rangle
& \qquad
\langle 0| {\cal O}_E^\dagger(\epsilon; T) \mathcal{E} {\cal O}_E (\epsilon'; T) |0\rangle \\
& \\
\langle 0| {\cal O}_E^\dagger(\epsilon'; T) \mathcal{E} {\cal O}_E (\epsilon; T) |0\rangle
& \qquad
\langle 0| {\cal O}_E^\dagger(\epsilon'; T) \mathcal{E} {\cal O}_E (\epsilon'; T) |0\rangle
\end{array}
\right),
\end{eqnarray}
where $\mathcal{E}$ is defined in (\ref{mathcale}). 
To be explicit let us  write out the elements in the row of the energy matrix. 
\begin{eqnarray}
&&\langle 0| {\cal O}_E^\dagger(\epsilon; T) \mathcal{E} {\cal O}_E (\epsilon; T) |0\rangle
= \frac{1}{\langle  {\cal O}_E (\epsilon; T) | {\cal O}_E (\epsilon, T) \rangle } \times 
\\ \nonumber
&&\qquad\qquad \qquad \int d^3 x e^{i Et} \lim_{x_1^+\rightarrow \infty} \frac{ x_1^+  - x_1^-}{4}
\int dx_1^- \langle \epsilon \cdot T (x)  T_{--}( x_1) \epsilon\cdot T(0) \rangle , 
\\ \nonumber
&&\langle  0|{\cal O}_E^\dagger(\epsilon; T) \mathcal{E} {\cal O}_E (\epsilon'; T)|0 \rangle
=( \langle {\cal O}_E (\epsilon'; T) | {\cal O}_E (\epsilon,' T) \rangle
\langle  {\cal O}_E (\epsilon; T) | {\cal O}_E (\epsilon, T) \rangle)^{-\frac{1}{2}}
\\ \nonumber
&& \qquad\qquad \qquad \int d^3 x e^{i Et} \lim_{x_1^+\rightarrow \infty} \frac{ x_1^+  - x_1^-}{4}
\int dx_1^- \langle \epsilon \cdot T (x)  T_{--}( x_1) \epsilon'\cdot T(0) \rangle.
\end{eqnarray}
The other entries of the matrix in (\ref{enmat}) are defined similarly. 
Essentially we need to evaluate ratios of $3$ point functions of the stress tensor, 
take the $x_1^+ \rightarrow \infty$ and then perform the integral over the space time
point of the last insertion of the stress tensor.

Let us examine the energy matrix corresponding to the charge excitations. 
We choose the two independent polarisations to be 
\begin{equation}
\epsilon^x = 1, \qquad \epsilon^{\prime y} = 1.
\end{equation}
The corresponding states are 
\begin{equation}
{\cal O}_{E} ( \epsilon; j) |0\rangle , \qquad {\cal O}_E( \epsilon';j) |0\rangle.
\end{equation}
The energy matrix which results from these two states are given by
\begin{eqnarray}\label{ejmat}
\hat E(j) = 
\left( \begin{array}{cc}
\langle 0| {\cal O}_E^\dagger (\epsilon; j) \mathcal{E} {\cal O}_E (\epsilon; j) |0\rangle
& \qquad
\langle 0| {\cal O}_E^\dagger(\epsilon; j) \mathcal{E} {\cal O}_E (\epsilon'; j) |0\rangle \\
& \\
\langle 0| {\cal O}_E^\dagger(\epsilon'; j) \mathcal{E} {\cal O}_E (\epsilon; j) |0\rangle
& \qquad
\langle 0| {\cal O}_E^\dagger(\epsilon'; j) \mathcal{E} {\cal O}_E (\epsilon'; j) |0\rangle
\end{array}
\right),
\end{eqnarray}
Again to be explicit, we write down the entries corresponding to the first row
\begin{eqnarray}
&&\langle 0| {\cal O}_E^\dagger(\epsilon; j) \mathcal{E} {\cal O}_E (\epsilon; j) |0\rangle
= \frac{1}{\langle  {\cal O}_E (\epsilon; j) | {\cal O}_E (\epsilon, j) \rangle } \times 
\\ \nonumber
&&\qquad\qquad \qquad \int d^3 x e^{i Et} \lim_{x_1^+\rightarrow \infty} \frac{ x_1^+  - x_1^-}{4}
\int dx_1^- \langle \epsilon \cdot j (x)  T_{--}( x_1) \epsilon\cdot T(0) \rangle , 
\\ \nonumber
&&\langle  0|{\cal O}_E^\dagger(\epsilon; j) \mathcal{E} {\cal O}_E (\epsilon'; j)|0 \rangle
=( \langle {\cal O}_E (\epsilon'; j) | {\cal O}_E (\epsilon,' j) \rangle
\langle  {\cal O}_E (\epsilon; j) | {\cal O}_E (\epsilon, j) \rangle)^{-\frac{1}{2}}
\\ \nonumber
&& \qquad\qquad \qquad \int d^3 x e^{i Et} \lim_{x_1^+\rightarrow \infty} \frac{ x_1^+  - x_1^-}{4}
\int dx_1^- \langle \epsilon \cdot j (x)  T_{--}( x_1) \epsilon'\cdot j(0) \rangle.
\end{eqnarray}

The condition of positivity of the energy observed by the calorimeter \citep{Hofman:2008ar} translates to demanding that the eigen values of the energy matrix 
(\ref{enmat}), (\ref{ejmat}) be positive. 
We will see that  only the parity even terms in (\ref{threepoint function})  contribute to the 
diagonal and  the parity odd term  
which contributes to the off diagonal entries 
of these matrices.

\section{Energy matrix for  charge excitations }\label{jjtcalculation}

The basic ingredient to evaluate the energy matrix for charge excitations
is the three point function of two $U(1)$ currents with 
a single insertion of the stress tensor. 
Including the parity odd tensor structures, this is given by
 \citep{Giombi:2011rz, Osborn:1993cr},
\begin{eqnarray}\label{jjt}
\langle j(x) T(x_1) j(0) \rangle &=& \frac{1}{|x_1-x|^3 |x_{1}|^3 |x|} \epsilon^\sigma_2 I_{\sigma}^\alpha (x-x_1) \epsilon^{\rho}_3 I_{\rho}^\beta (-x_1) \epsilon_1^{\mu \nu} t_{\mu \nu \alpha \beta} (X) \nonumber\\
&& + p_j\frac{Q_1^2S_1+2P_2^2S_3+2P_3^2S_2}{|x_1-x||x||-x_1|},
\end{eqnarray} 
where the first line is the usual parity even contribution while the second line is the parity odd contribution. 
We use the conventions of \cite{Giombi:2011kc} for the 
 normalisation of the parity odd tensor structure.
 
These tensor structures are listed
 below. 
\begin{eqnarray}
t_{\mu \nu \alpha \beta}(X) &=&(-\frac{2 c}{3}+2 e) h^1_{\mu \nu}(\hat{X}) \eta_{\alpha \beta} + (3 e) h^1_{\mu \nu}(\hat{X}) h^1_{\alpha \beta} + c h^2_{\mu \nu \alpha \beta}(\hat{X}) + e h^3_{\mu \nu \alpha \beta},\nonumber\\
Q_1^2 &=& \epsilon^\mu_1 \epsilon^\nu_1 \left(\frac{x_{1\mu}}{x_1^2}-\frac{x_{1\mu}-x_\mu}{(x_1-x)^2}\right)\left(\frac{x_{1\nu}}{x_1^2}-\frac{x_{1\nu}-x_\nu}{(x_1-x)^2}\right),\nonumber\\
P_2^2&=& -\frac{\epsilon^{ \mu}_1 \epsilon^\nu_1 I_{\mu\nu}(x_1)}{2x_1^2},\nonumber\\
P_3^2&=& -\frac{\epsilon^\mu_1\epsilon^\nu_2 I_{\mu\nu}(x_1-x)}{2(x_1-x)^2},\nonumber\\
S_1 &=& \frac{1}{4|x_1-x||x|^3|-x_1|}\left( \varepsilon^{\mu\nu}_{\phantom{\mu\nu}\rho} x_{\mu} (x_1-x)_{\nu} \epsilon^\rho_2 \epsilon^\alpha_3 x_{\alpha} - 
\frac{\varepsilon^\mu_{\phantom{\mu}\nu\rho}}{2}\left( |x_1-x|^2x_{\mu} + |x|^2 (x_1-x)_\mu \right) \epsilon_2^\nu \epsilon_3^\rho \right),\nonumber\\
S_2 &=& \frac{1}{4|x_1-x||x||-x_1|^3}\left( \varepsilon^{\mu\nu}_{\phantom{\mu\nu}\rho} (x_{1\mu}) x_{\nu} \epsilon^\rho_3 \epsilon^\alpha_1 x_{1\alpha} - 
\frac{\varepsilon^\mu_{\phantom{\mu}\nu\rho}}{2}\left( -|x|^2x_{1\mu} + |x_1|^2 x_{\mu} \right) \epsilon_3^\nu \epsilon_1^\rho \right),\nonumber\\
S_3 &=& \frac{1}{4|x_1-x|^3|x||-x_1|}\left( \varepsilon^{\mu\nu}_{\phantom{\mu\nu}\rho}(x_1-x)_{\mu}(-x_{1\nu}) \epsilon^\rho_1 \epsilon^\alpha_2 (x_1-x_\alpha) \right.  \nonumber\\
&&\phantom{\frac{4}{|x-x_1|^3|x_1||-x|}} \left.- \frac{\varepsilon^\mu_{\phantom{\mu}\nu\rho}}{2}\left( |x|^2(x-x_1)_\mu + |x-x_1|^2 (-x_\mu) \right) \epsilon_1^\nu \epsilon_2^\rho \right),\nonumber\\
\end{eqnarray}
where,
\begin{eqnarray}
\hat X &=& \frac{x-x_1}{|x-x_1|^2}+\frac{x_1}{|x_1|^2} ,\nonumber\\
I_{\alpha \beta} (x) &=& \eta_{\alpha \beta} -\frac{2 x_\alpha x_\beta}{x^2},\nonumber\\
h^1_{\mu \nu}(\hat{x}) &=& \frac{x_\mu x\nu}{x^2}-\frac{1}{3} \eta_{\mu \nu},\nonumber\\\
h^2_{\mu \nu \sigma \rho}(\hat{x})&=& \frac{x_\mu x_\sigma}{x^2}\eta_{\nu \rho}+ \left( \mu \leftrightarrow \nu, \rho \leftrightarrow \sigma \right) -\frac{4}{3} \frac{x_\mu x_\nu}{x^2}\eta_{\sigma\rho}-\frac{4}{3} \frac{x_\sigma x_\rho}{x^2}\eta_{\mu \nu} + \frac{3}{16} \eta_{\mu \nu} \eta_{\sigma \rho},  \nonumber\\
h^3{\mu \nu \sigma \rho} &=& \eta_{\mu \sigma} \eta_{\nu \rho} +\eta_{\mu \rho} \eta_{\nu \sigma} -\frac{2}{3} \eta_{\mu \nu} \eta_{\sigma \rho},\nonumber\\
c &=& \frac{3 (2 n^j_f + n^j_s)}{256 \pi ^3}, \qquad
e = \frac{3 n^j_s}{256 \pi ^3}. \nonumber\\
\end{eqnarray}   

For normalising the excited states, we also need the two point function 
of currents which is given by 
\begin{eqnarray}\label{2ptjj}
\langle j_\mu ( x) j_\nu\rangle = &=& \frac{C_V}{x^{4}}I_{\mu \nu} (x),
\end{eqnarray}
with 
\begin{equation}
C_V = \frac{8}{3} \pi  (c+e).
\end{equation}

\subsection{Parity even contribution}

That contribution to the energy deposited in the conformal collider  due to 
charge excitations from the 
parity even part has been obtained before for arbitrary dimensions in  \cite{Chowdhury:2012km}. 
Here we repeat this analysis as a cross check as well as to fix our 
conventions. 
Its easy to see that 
diagonal terms in the energy matrix (\ref{jjt}) result only from the 
parity even terms  in the three point function (\ref{ejmat}).
Let us first choose the 
the  polarisations  to be given by 
\begin{eqnarray}\label{jjtevenpol1}
\epsilon_2^x=\epsilon_3^x = \epsilon_{1}^{--} = 1.
\end{eqnarray}
This corresponds to   the first entry of the energy matrix and is given by 
the ratio
\begin{eqnarray} \label{ratioj}
\hat E(j)_{11}
&=& \frac{g^1_j(E)}{g^2_j(E)},
\end{eqnarray}  
\begin{eqnarray}
g^1_j(E) &=& \int d^3x e^{iEt} \lim_{x_1^+ \rightarrow \infty}\frac{x_1^+ - x_1^-}{4}\int_{-\infty}^{\infty} {dx^-} \langle j_x(x) T_{--}(x_1) j_x(0)  \rangle,\nonumber\\
\end{eqnarray}
and
\begin{eqnarray}
g_j^2(E) = \int d^3 x e^{- i Et} \langle j_x ( x) j_x(0) \rangle.
\end{eqnarray}
We look at the individual contributions to $g^1_j(E)$. On substituting 
the expression for the three point function in (\ref{jjt}) we obtain 
four terms
\begin{equation}
g^1_j(E) = I'_1+I'_2+I'_3+I'_4,
\end{equation}
\begin{eqnarray}
I'_1 &=&  \int d^3x e^{iEt} \lim_{x_1^+ \rightarrow \infty}\frac{x_1^+ - x_1^-}{4}\int_{-\infty}^{\infty} {dx^-} (-\frac{2 c}{3}+2 e) I_x^\alpha(x-x_1) I_x^\beta(-x_1) h^1_{--}(\hat{X}) \eta_{\alpha \beta}. \nonumber\\
\end{eqnarray}
 Let us describe the steps involved 
 in evaluating the limit and the integral. First the limit $x_1^+ \rightarrow \infty$ is taken
 which results in 
\begin{eqnarray}
I'_1 &=& (-\frac{2 c}{3}+2 e) \int d^3x e^{iEt} \int_{-\infty}^{\infty} {dx^-}  \frac{(x^-)^2}{16 (x^--x_1^--i\epsilon)^{5/2} (-x_1^-+i\epsilon)^{5/2} \left(x^2-x^- x^+\right)^{3/2}} .\nonumber\\
\end{eqnarray}
We follow the $i\epsilon$ prescription introduced by 
 \citep{Hofman:2008ar, Buchel:2009sk} to evaluate the integral.
  Operators are assigned 
a negative imaginary part to time depending on their position in the correlation function. 
Operators to the left are assigned a larger negative imaginary part than the operators 
to the right i.e, $t_1\rightarrow t_1 -i\epsilon$, $t \rightarrow t -2i\epsilon$. Hence the light-cone coordinates change
$x^{\pm}_1\rightarrow x^{\pm}_1 -i\epsilon$, $x^{\pm} \rightarrow x^{\pm} -2i\epsilon$. The integral over
$x_1^-$ is then performed by using  the  integral in (\ref{F3}).
The integrals over the other directions are then 
best performed by integrating over $x$ direction, followed by integrating the 
light cone directions.  
\begin{eqnarray}
I'_1 &=& \frac{1}{4} (-\frac{2 c}{3}+2 e) \int dx^+ dx^- e^{\frac{iEx^+}{2}}e^{\frac{iEx^-}{2}} \frac{16 
}{3 (x^--2i\epsilon)^2 {x^- x^+}},\nonumber\\
&=&\frac{2}{3}  \left( {-\frac{2 c}{3}+2 e}\right)  E^2 \pi ^2.
\end{eqnarray}  
where we have used \ref{F3} and \ref{F4}. 
Following the same method we obtain 
\begin{eqnarray}
I'_2 &=&  \int d^3x e^{iEt} \lim_{x_1^+ \rightarrow \infty}\frac{x_1^+ - x_1^-}{2}\int_{-\infty}^{\infty} \frac{dx^-}{2} (3 e) I_x^\alpha(x-x_1) I_x^\beta(-x_1) h^1_{--}(\hat{X}) h^1_{\alpha \beta},\nonumber\\
&=& (3 e)\int d^3x e^{iEt} \int_{-\infty}^{\infty} \frac{dx^-}{2} \frac{(x^-)^2 \left(2 x^2+x^- x^+\right)}{24 (x^--x_1^-)^{5/2} (-x_1^-)^{5/2} \left(x^2-x^- x^+\right)^{5/2}},\nonumber\\
&=& 0.
\end{eqnarray} 
\begin{eqnarray}
I'_3 &=&  \int d^3x e^{iEt} \lim_{x_1^+ \rightarrow \infty}\frac{x_1^+ - x_1^-}{4}\int_{-\infty}^{\infty} 
dx^- c I_x^\alpha(x-x_1) I_x^\beta(-x_1) h^2_{-- \alpha \beta}(\hat{X}), \nonumber\\
&=& - c\int d^3x e^{iEt}\int_{-\infty}^{\infty} \frac{dx^-}{12}  -\frac{(x^-)^2}{(x^--x_1^-)^{5/2} (-x_1^-)^{5/2} \left(x^2-x^- x^+\right)^{3/2}}, \nonumber\\
&=& -\frac{c}{4}\int dx^+ dx^- e^{\frac{iEx^+}{2}}e^{\frac{iEx^-}{2}}(\frac{64}{9 (x^--2i\epsilon)^2 \sqrt{(x^- x^+)^2+ i\epsilon x^-x^+}}),\nonumber\\
&=& \frac{c}{2}\left(\frac{-16}{9} E^2 \pi ^2\right).
\end{eqnarray}    
  \begin{eqnarray}
  I'_4 &=&  \int d^3x e^{iEt} \lim_{x_1^+ \rightarrow \infty}\frac{x_1^+ - x_1^-}{2}\int_{-\infty}^{\infty} \frac{dx^-}{2} e I_x^\alpha(x-x_1) I_x^\beta(-x_1) h^3_{-- \alpha \beta}, \nonumber\\
  &=& 0.
\end{eqnarray}   
Putting all this together the result for the numerator in (\ref{ratioj}) is given by 
\begin{eqnarray}
g^1_j(E) &=& \frac{4}{3} E^2 \pi ^2 (e- c).
\end{eqnarray}
Now the denominator in (\ref{ratioj}) is defined by
\begin{eqnarray}
g^2_j(E) &=& \int d^3 x e^{- i Et} \langle j_x ( x) j_x(0) \rangle,\nonumber\\
&=& -\frac{8}{3} \pi  (c+e) \int d^3 x e^{- i Et} \frac{x^2+x^- x^+}{\left(x^2-x^- x^+\right)^3}.
\end{eqnarray}
The calculation proceeds similarly as before, with the spatial integrals being performed first, followed by the light cone directions using the $i\epsilon$ prescription.
\begin{eqnarray}
g^2_j(E) &=& -\frac{8}{3} \pi  (c+e) \left(\frac{1}{2}\int dx^+ dx^- e^{\frac{iEx^+}{2}}e^{\frac{iEx^-}{2}} \frac{i \pi }{4 \left( x ^- -2i\epsilon \right)^{3/2} \left(x^+ -2i\epsilon \right)^{3/2}}\right), \nonumber\\
&=&- \frac{8}{3}  E \pi ^3 (c+e).
\end{eqnarray}
Therefore we obtain 
\begin{eqnarray}\label{eje11}
\hat E(j)_{11} &=& \frac{(c-e) E}{2 (c+e) \pi },\nonumber\\
&=& \frac{E}{4\pi}(1-\frac{a_2}{2}),
\end{eqnarray}       
where,
\begin{equation}\label{a2def}
a_2 = \frac{2(3e-c)}{(e+c)} = -\frac{2 (n^j_f-n^j_s)}{(n^j_f+n^j_s)}.
\end{equation}

Let us examine the second diagonal element of the charge matrix. 
For this we choose the polarisations to be given by 
\begin{eqnarray}
\epsilon_2^y=\epsilon_3^y = \epsilon_{1}^{--} = 1.
\end{eqnarray}
Proceeding  identically we obtain 
\begin{eqnarray}\label{eje22}
\hat E(j)_{22} &=& \frac{e }{c  +e  } \frac{E}{\pi},\nonumber\\
&=& \frac{E}{4\pi}(1+\frac{a_2}{2}).
\end{eqnarray} 
If we restrict the class of theories to be parity preserving, then  from (\ref{eje11}) and 
(\ref{eje22}) 
we obtain the constraint
\begin{equation}
|a_2 | \leq 2.
\end{equation}
This agrees with the results of 
 \citep{Chowdhury:2012km}.

\subsection{Parity odd contribution}

It can be seen that the off diagonal contribution is entirely due to the 
parity odd terms in the three point function (\ref{jjt}). 
Let us first examine the $(12)$ element of the energy matrix. 
For this we choose the polarizations to be given by 
\begin{eqnarray}
\epsilon_2^{x} =\epsilon_3^{y}=1 .
\end{eqnarray}
Again this can be written as a ratio
\begin{eqnarray}\label{ratio2}
\hat E(j)_{12} &=& \frac{f^1_j(E)}{f^2_j(E)}.
\end{eqnarray}
From the structure of the parity odd term, the numerator naturally 
breaks up into 3 parts which are defined as follows.
\begin{eqnarray}\label{numf}
f^1_j(E) &=& \int d^3x e^{iEt}\; \lim_{x_1^+ \rightarrow \infty}\frac{x_1^+ - x_1^-}{2}\int_{-\infty}^{\infty} \frac{dx_1^-}{2}  j_x (x) T_{--}(x_1)j_y(0),\nonumber\\
&=& \mathcal{I}^p_1+\mathcal{I}^p_2+\mathcal{I}^p_3,
\end{eqnarray}
where,
\begin{eqnarray}
\mathcal{I}^p_n &=& \int d^3x e^{iEt}\; \lim_{x_1^+ \rightarrow \infty}\frac{x_1^+ - x_1^-}{2}\int_{-\infty}^{\infty} \frac{dx_1^-}{2} I^p_n .
\end{eqnarray}
Now let us evaluate each of the integrals following the same methods introduced 
earlier.  The first integrand is given by
\begin{eqnarray}
I^p_1 &=& p_j\frac{Q_1^2S_1}{|x_1-x||x||-x_1|}_{\epsilon_1^{--} =\epsilon_2^{x}= \epsilon_3^{y} =1} ,\nonumber\\
&=& \frac{16p_j}{64|x_1-x|^2|x|^4|-x_1|^2}\left(\frac{x_{1-}}{x_1^2}-\frac{(x_1-x)_-}{(x_1-x)^2}\right)^2\nonumber\\
&&\left( \varepsilon^{\mu\nu}_{\phantom{\mu\nu}x} x_{\mu} (x_1-x)_{\nu} (x_{+}-x_-) -
 \frac{\varepsilon^\mu_{\phantom{\mu}x+}-\varepsilon^\mu_{\phantom{\mu}x-}}{2}\left( |x_1-x|^2x_{\mu} + |x|^2 (x_1-x)_\mu \right)\right),\nonumber\\ 
\end{eqnarray}
where the $\varepsilon$ tensor is given by 
\begin{equation}
\varepsilon_{+-x} = \frac{1}{2}.
\end{equation}
Taking the $x_1^+\rightarrow \infty$ limit and then performing the integral
we obtain
\begin{eqnarray}
\mathcal{I}^p_1 &=& \int d^3x e^{iEt}\; \lim_{x_1^+ \rightarrow \infty}\frac{x_1^+ - x_1^-}{2}\int_{-\infty}^{\infty} \frac{dx_1^-}{2} I^p_1, \nonumber\\
&=&p_j\int d^3x e^{iEt}\;\int_{-\infty}^{\infty} \frac{dx_1^-}{4} \frac{ (x^-)^2 \left(x^2+x^- (x^+-x_1^-)-x_1^- x^+\right)}{64 (x_1^-)^3 (x^--x_1^-)^3 \left(x^2-x^- x^+\right)^2},\nonumber\\
&=& \frac{p_j}{128}\int d^3x e^{iEt} 
\left[{\frac{6 i \pi    \left(x^2+x^- x^+\right)}{(x^-)^3}+\frac{3 i \pi   (-x^--x^+)}{(x^-)^2}}
\right]\left[\frac{1}{\left(x^2-x^- x^p\right)^2}\right],\nonumber\\
&=&-\frac{p_j}{128}\frac{1}{4}\int dx^+ dx^- e^{\frac{iEx^+}{2}}e^{\frac{iEx^-}{2}} \left(\frac{3 \pi ^2 }{(x^--2i\epsilon)^{5/2} (x^+-2i\epsilon)^{3/2}}+\frac{3 \pi ^2  }{(x^--2i\epsilon)^{7/2} \sqrt{x^+-2i\epsilon}}\right),\nonumber\\
&=& -\frac{3 p_j}{160} E^2 \pi ^3 .
\end{eqnarray}
The  integrand for the second term in (\ref{numf}) is given by 
\begin{eqnarray}
&&I^p_2 = p_j\frac{2P_2^2S_3}{|x_1-x||x||-x_1|}_{\epsilon_1^{--} =\epsilon_2^{x}= \epsilon_3^{y} =1}, \nonumber\\
&&= \frac{-p_j}{4|x_1-x|^4|x|^2|-x_1|^4}I_{y-}(x_1)\\
&&\left( \varepsilon^{\mu\nu}_{\phantom{\mu\nu}-} (x_1-x)_{\mu} (-x_1)_{\nu} (x_1-x)_x - \frac{\varepsilon^\mu_{\phantom{\mu}-x}}{2}\left( |-x_1|^2(x_1-x)_{\mu} + |x_1-x|^2 (-x_1)_\mu \right)\right).\nonumber
\end{eqnarray}
Taking the limit and performing the integrals we obtain
\begin{eqnarray}
\mathcal{I}^p_2 &=& \int d^3x e^{iEt}\; \lim_{x_1^+ \rightarrow \infty}\frac{x_1^+ - x_1^-}{4}\int_{-\infty}^{\infty} {dx_1^-} I^p_2, \nonumber\\
&=&\frac{p_j}{128} \int d^3x e^{iEt}\;\int_{-\infty}^{\infty} {dx_1^-}\frac{  x^-}{(x_1^-)^3 (x^--x_1^-)^2 \left(x^2-x^- x^+\right)} ,\nonumber\\
&=& \frac{p_j}{256}\int d^3x e^{iEt} \frac{12 i \pi   }{(x^-)^3 \left(x^2-x^- x^+\right)} ,\nonumber\\
&=&- \frac{p_j}{80}  E^2 \pi ^3 .
\end{eqnarray}
Finally the integrand for the last term in (\ref{numf}) is given by 
\begin{eqnarray}
I^p_3 &=& p_j\frac{2P_3^2S_2}{|x_1-x||x||-x_1|}_{\epsilon_1^{--} =\epsilon_2^{x}= \epsilon_3^{y} =1}, \nonumber\\
&=& \frac{-16p_j}{64|x_1-x|^4|x|^2|-x_1|^4}I_{-x}(x_1-x)\nonumber\\
&&\left( (\varepsilon^{\mu\nu}_{\phantom{\mu\nu}+}-\varepsilon^{\mu\nu}_{\phantom{\mu\nu}-}) (x_1)_{\mu} x_{\nu} (x_1)_- - \frac{\varepsilon^\mu_{\phantom{\mu}+-}}{2}\left( -|x|^2 x_{1\mu} + |x_1|^2 x_\mu \right)\right),\nonumber\\
\end{eqnarray}
Evaluating the limit and the integrals we obtain
\begin{eqnarray}
\mathcal{I}^p_3 &=& \int d^3x e^{iEt}\; \lim_{x_1^+ \rightarrow \infty}\frac{x_1^+ - x_1^-}{2}\int_{-\infty}^{\infty} \frac{dx_1^-}{2} I^p_3, \nonumber\\
&=& 0.
\end{eqnarray}
 Putting all the terms for the numerator together we get
\begin{eqnarray}
f_1(E) &=& \mathcal{I}^{p}_{1}+\mathcal{I}^{p}_{2}+\mathcal{I}^{p}_{3},\nonumber\\
&=& -\frac{E^2 \pi ^3  p_j}{32}.\nonumber
\end{eqnarray}
The denominator in (\ref{ratio2})  is given by
\begin{eqnarray}
f_2(E) &=& -\frac{8}{3}  E\pi ^3 (c+e).
\end{eqnarray}
Therefore the $(12)$ element of the energy matrix is given by 
\begin{eqnarray}
\hat E(j)_{12}  
&=&\frac{ 3p_j  E}{256(c+e)},\nonumber\\
&=& \frac{E}{8\pi} \alpha_j,
\end{eqnarray}
where $\alpha_j$ is defined as 
\begin{eqnarray}\label{alphajdef}
\alpha_j &\equiv& \frac{ 3 p_j  \pi}{32(c+e)}=\frac{4 \pi ^4 p_j}{(n^j_f+n^j_s)}.
\end{eqnarray}
Let us now examine the second off diagonal element in the energy matrix. 
To extract this component  we choose the polarisations to be given by 
\begin{eqnarray}
\epsilon_2^y=\epsilon_3^x=1.
\end{eqnarray}
Again this element is given by the ratio
\begin{eqnarray}
\hat E(j)_{21}  &=& \frac{f'^1_j(E)}{f^2_j(E)}.
\end{eqnarray}
By evaluating the numerator explicitly using the same methods it can be seen that 
\begin{equation}
f'^1_j(E) = f^1_j(E).
\end{equation}
Therefore the two off diagonal elements are identical.
This is consistent with the fact that the energy matrix must be symmetric and is in fact a
cross check for our calculations.

To summarise, the energy matrix for charge excitations is given by 
 \begin{eqnarray} \label{ejmat1}
 \hat E(j) = \begin{pmatrix}
\frac{E}{4\pi}(1-\frac{a_2}{2})&&\frac{E}{8\pi} \alpha_j\\
 \\
\frac{E}{8\pi} \alpha_j && \frac{E}{4\pi}(1+\frac{a_2}{2})\\
 \end{pmatrix},
\end{eqnarray}
where
\begin{eqnarray}\label{entryjmat}
a_2 &=& \frac{2(3e-c)}{(e+c)} = -\frac{2 (n^j_f-n^j_s)}{(n^j_f+n^j_s)} ,\nonumber\\
\alpha_j &=& \frac{ 3 p_j  \pi}{32(c+e)} = \frac{4 \pi ^4 p_j}{(n^j_f+n^j_s)}.
\end{eqnarray}
The condition that the energy observed at the calorimeter is positive leads to the 
fact that the eigen values of of the energy matrix are positive.
The trace of the  matrix in (\ref{ejmat1}) is positive, this implies that the determinant
is positive which leads to the condition
\begin{equation}
a_2^2 + \alpha_j^2 \leq 4.
\end{equation}
This region is a disc of radius $2$ centered at the origin in the $a_2, \alpha_j$ plane. 

\section{Energy matrix for stress tensor excitations}\label{tttcalculation}

In this section we evaluate the energy matrix corresponding to 
stress tensor excitations $\hat E(T)$  defined in (\ref{enmat}). 
Here the basic ingredient is the three point function of the stress energy 
tensor including the parity odd term. 
The parity even tensor structures  in the three point function was found earlier 
by \citep{Osborn:1993cr,Erdmenger:1996yc} while the parity 
odd contribution was written down by \cite{Giombi:2011rz}.
Combining both these contributions the three point function of 
the stress tensor for a conformal field theory in $d=3$ is given by 
\begin{eqnarray}\label{ttt}
\langle T(x)T(x_1)T(0) \rangle &=&\frac{\epsilon_1^{\mu \nu} \mathcal{I}^T_{\mu \nu, \mu' \nu'}(x)\epsilon^{\sigma \rho}_2 \mathcal{I}^T_{\sigma \rho, \sigma' \rho'}(x_1) \epsilon^{\alpha \beta}_3 t^{\mu' \nu' \sigma' \rho'}_{\qquad \alpha \beta}}{x^6 x_1^6} + \nonumber\\
&&p_T\frac{(P_1^2Q_1^2+5P_2^2P_3^2)S_1+(P_2^2Q_2^2+5P_3^2P_1^2)S_2+(P_3^2Q_3^2+5P_3^2P_1^2)S_3}{|x-x_1||x_1||-x|}.\nonumber\\
\end{eqnarray}
Here the first line denotes the  parity even contribution to the three point function and the parity odd contribution is given in the second line. 
We will choose  the normalisation of the parity odd tensor structure as 
given in \cite{Giombi:2011kc}. 
The parity even tensor structures are given by 
\begin{eqnarray}\label{parite}
t_{\mu \nu \sigma \rho \alpha \beta} &=& \mathcal{A} \mathcal{E}^T_{\mu \nu, \epsilon \eta } {\mathcal{E}^T}_{\sigma \rho,\;\lambda}^{\;\;\phantom{\rho}\eta\;} {\mathcal{E}^T}_{\alpha \beta,}^{\phantom{\alpha}\phantom{\beta}\lambda \epsilon} \frac{1}{(Z^2)^{\frac{3}{2}}} \\
&&+(\mathcal{B}-2\mathcal{A}) \mathcal{E}^T_{\alpha \beta, \epsilon \eta}{\mathcal{E}^T}_{\sigma \rho,\;\kappa}^{\;\;\phantom{\rho}\eta\;} {\mathcal{E}^T}_{\mu\nu, \lambda}^{\phantom{\mu\nu, \lambda} \epsilon} \frac{Z^\kappa Z^\lambda}{(Z^2)^{\frac{5}{2}}} \nonumber\\
&& - \mathcal{B}\left(\mathcal{E}^T_{\mu \nu, \epsilon \eta } {\mathcal{E}^T}_{\sigma \rho,\phantom{\eta}\kappa}^{\;\;\phantom{\sigma \rho}\eta\;}{\mathcal{E}^T}_{\alpha \beta,\lambda}^{\phantom{\alpha}\phantom{\beta}\phantom{\lambda} \epsilon} + \left( \mu\nu \right) \leftrightarrow \left(\sigma \rho \right)\right)\frac{Z^\kappa Z^\lambda}{(Z^2)^{\frac{5}{2}}}\nonumber\\
&& + \mathcal{C}\left(\mathcal{E}^T_{\mu \nu, \sigma \rho}(\frac{Z_{\alpha}Z_{\beta}}{Z^2}-\frac{1}{3} \eta_{\alpha \beta}) + \mathcal{E}^T_{\sigma \rho,\alpha \beta}(\frac{Z_{\mu}Z_{\nu}}{Z^2}-\frac{1}{3} \eta_{\mu \nu}) \right. \nonumber\\
&& \left. \qquad \mathcal{E}^T_{\alpha \beta,\mu \nu}(\frac{Z_{\sigma}Z_{\rho}}{Z^2}-\frac{1}{3} \eta_{ \sigma \rho}) \right)\frac{1}{(Z^2)^{\frac{3}{2}}}\nonumber\\
&&+ (\mathcal{D}-4\mathcal{C})\left(\mathcal{E}^T_{\mu \nu, \epsilon \kappa}{\mathcal{E}^T}_{\sigma \rho, \phantom{\epsilon} \lambda}^{\phantom{\sigma \rho,}\epsilon}  \right)\left(\frac{Z_{\alpha}Z_{\beta}}{Z^2}-\frac{1}{3} \eta_{\alpha \beta} \right)\frac{Z^\kappa Z^\lambda}{(Z^2)^{\frac{5}{2}}}\nonumber\\
&& - (\mathcal{D}-2\mathcal{B}) \left(\mathcal{E}^T_{\sigma \rho,\epsilon \kappa} {\mathcal{E}^T}_{\alpha \beta, \phantom{\epsilon} \lambda}^{\phantom{\sigma \rho,}\epsilon}\left(\frac{Z_{\mu}Z_{\nu}}{Z^2}-\frac{1}{3} \eta_{\mu \nu}\right)+ \left( \mu\nu \right) \leftrightarrow \left(\sigma \rho \right)\right)\frac{Z^\kappa Z^\lambda}{(Z^2)^{\frac{5}{2}}}\nonumber\\
&& + (\mathcal{E} + 4\mathcal{C}- 2 \mathcal{D})\left(\frac{Z_{\mu}Z_{\nu}}{Z^2}-\frac{1}{3} \eta_{\mu \nu}\right)\left(\frac{Z_{\alpha}Z_{\beta}}{Z^2}-\frac{1}{3} \eta_{\alpha \beta} \right) \left(\frac{Z_{\sigma}Z_{\rho}}{Z^2}-\frac{1}{3} \eta_{ \sigma \rho} \right)\frac{1}{Z^{\frac{3}{2}}}, \nonumber
\end{eqnarray}
where tensors  involved  and  the theory dependent parameters given by
\begin{eqnarray}
\mathcal{E}^T_{\mu\nu, \alpha \beta} &=& \frac{1}{2}(\eta_{\mu \alpha} \eta_{\nu \beta} + \eta_{\mu \beta} \eta_{\nu \alpha})- \frac{1}{3} \eta_{\alpha \beta} \eta_{\mu \nu},\\
I_{\alpha \beta} (x) &=& \eta_{\alpha \beta} -\frac{2 x_\alpha x_\beta}{x^2},\nonumber\\
\mathcal{I}^T_{\mu \nu, \alpha \beta}(x) &=& I_{\mu \mu'} (x) I_{\nu \nu'} (x) {\mathcal{E}^{T}}^{\mu' \nu',}_{\phantom{\mu' \nu'}\alpha \beta},\nonumber\\
\mathcal{D} &=& \frac{5 \mathcal{A}}{2}+\frac{5 \mathcal{B}}{2}-6 \mathcal{C}, \qquad \mathcal{E} = 5 \mathcal{A}+\frac{3}{4} (9\mathcal{B}-26 \mathcal{C}),\nonumber\\
\mathcal{A} &=& \frac{27 n_s^T}{512 \pi ^3}, \qquad \mathcal{B} = -\frac{9 (4 n_f^T + 3n_s^T )}{512 \pi ^3}, \qquad \mathcal{C} = -\frac{9 (8 n_f^T + n_s^T)}{2048 \pi ^3}. \nonumber
\end{eqnarray}
The parity odd tensor structures are given by 
\begin{eqnarray}
P_1^2&=& \frac{-\epsilon^\mu_2\epsilon^\nu_3 I_{\mu\nu}(x_1)}{2x_1^2}, \qquad
P_2^2= \frac{-\epsilon^\mu_3\epsilon^\nu_1 I_{\mu\nu}(x)}{2x^2},\\
P_3^2&=& \frac{-\epsilon^\mu_1\epsilon^\nu_2 I_{\mu\nu}(x-x_1)}{2(x-x_1)^2}, \nonumber \\
Q'_{1\alpha} &=& \left( \frac{x_\alpha}{x^2} - \frac{(x-x_1)_\alpha}{(x-x_1)^2}\right), \qquad
Q'_{2\alpha} = \left( -\frac{x_{1\alpha}}{x_1^2} +\frac{(x_1-x)_\alpha}{(x_1-x)^2}\right),\nonumber\\
Q'_{3\alpha} &=& \left( \frac{x_{1\alpha}}{x_1^2} - \frac{x_\alpha}{x^2}\right),\nonumber\\
Q_1^2 &=& \epsilon_1^\alpha \epsilon_1^\beta Q'_{1\alpha}Q'_{1\beta}, \nonumber \qquad
Q_2^2 = \epsilon_2^\alpha \epsilon_2^\beta Q'_{2\alpha}Q'_{2\beta}, \nonumber\\
Q_3^2 &=& \epsilon_3^\alpha \epsilon_3^\beta Q'_{3\alpha}Q'_{3\beta}, \nonumber
\end{eqnarray}
and 
\begin{eqnarray}
S_1 &=& \frac{1}{4|x-x_1||x_1|^3|-x|}\left( \varepsilon^{\mu\nu}_{\phantom{\mu\nu}\rho} x_{1\mu} (x-x_1)_{\nu} \epsilon^\rho_2 \epsilon^\alpha_3 x_{1\alpha} - \frac{\varepsilon^\mu_{\phantom{\mu}\nu\rho}}{2}\left( |x-x_1|^2x_{1\mu} + |x_1|^2 (x-x_1)_\mu \right) \epsilon_2^\nu \epsilon_3^\rho \right),\nonumber\\
S_2 &=& \frac{1}{4|x-x_1||x_1||-x|^3}\left( \varepsilon^{\mu\nu}_{\phantom{\mu\nu}\rho} (-x_{\mu}) x_{1\nu} \epsilon^\rho_3 \epsilon^\alpha_1 (-x_{\alpha}) - \frac{\varepsilon^\mu_{\phantom{\mu}\nu\rho}}{2}\left( |x_1|^2(-x_{\mu}) + |x|^2 x_{1\mu} \right) \epsilon_3^\nu \epsilon_1^\rho \right),\nonumber\\
S_3 &=& \frac{1}{4|x-x_1|^3|x_1||-x|}\left( \varepsilon^{\mu\nu}_{\phantom{\mu\nu}\rho}(x-x_1)_{\mu}(-x_\nu) \epsilon^\rho_1 \epsilon^\alpha_2 (-x_\alpha) \right.  \nonumber\\
&&\phantom{\frac{4}{|x-x_1|^3|x_1||-x|}} \left.- \frac{\varepsilon^\mu_{\phantom{\mu}\nu\rho}}{2}\left( |x|^2(x-x_1)_\mu + |x-x_1|^2 (-x_\mu) \right) \epsilon_1^\nu \epsilon_2^\rho \right).\nonumber\\
\end{eqnarray}

For obtaining the normalized states created by the stress tensor we also
need its two point function which is given by
\begin{eqnarray}\label{2pttt}
\langle T_{\mu \nu}(x) T_{\sigma \rho}(0) &=& \frac{C_T}{x^6}\mathcal{I}^T_{\mu \nu, \sigma \rho}(x), 
\end{eqnarray}
with 
\begin{equation}
C_T = \frac{2}{15} \pi  (10 \mathcal{A}-2 \mathcal{B}-16 \mathcal{C}).
\end{equation}

\subsection{Parity even contribution}

Though the derivation of the parity even contribution to the energy at the 
conformal collider in $d=3$ can be obtained from the results of
\cite{Buchel:2009sk} we repeat the analysis as a check.
It can be seen that it is only 
the parity even terms in (\ref{ttt}) that contribute to the diagonal terms of the 
energy matrix $\hat E(T)$. 
From (\ref{parite}) we write  the parity even part of the three point function 
as a contribution from 7 terms.

\begin{eqnarray}\label{tttevenb}
\langle T_{\mu\nu} (x) T_{\sigma \rho} (x_1) T_{\alpha \beta} (0) \rangle_{\rm even},
 &=& I_7 + I_6 + I_5 + I_4 + I_3 + I_2 + I_1 .
\end{eqnarray}

Let us proceed to evaluate the $(11)$ element of $\hat E(T)$ given in 
(\ref{enmat}).  For this we choose the polarization to be given by, 
\begin{eqnarray}
&&\epsilon_1^{xy} = \epsilon_1^{yx} = \frac{1}{2}, \qquad
\epsilon_3^{xy} = \epsilon_3^{yx} = \frac{1}{2}, \\ \nonumber
&&\epsilon_2^{--} = 1 .
\end{eqnarray}
The matrix element can be written as the ratio
\begin{eqnarray} \label{ratiott}
\hat E(T)_{11}
&=& \frac{g^1_T(E)}{g^2_T(E)}.
\end{eqnarray}
The denominator in  (\ref{ratiott})  is the norm which depends on the 
two point function of the stress tensor. We will first examine the numerator. 
From the  break up the three point function in (\ref{tttevenb}) we have,
\begin{eqnarray} \label{ttterms}
g^1_T(E) &=& \int d^3x e^{iEt}\; \lim_{x_1^+ \rightarrow \infty}\frac{x_1^+ - x_1^-}{4}\int_{-\infty}^{\infty} {dx_1^-}\langle  T_{xy}(x) T_{--}(x_1)T_{xy}(0) \rangle, \nonumber\\
&=& \mathcal{I}_7 + \mathcal{I}_6 + \mathcal{I}_5 +\mathcal{I}_4 +\mathcal{I}_3 +\mathcal{I}_2 + \mathcal{I}_1.
\end{eqnarray}
Performing the limit and the integrals in each of the 7 terms is tedious but straight forward.
The procedure is identical to that carried out for the charge excitations 
in the previous section. 
The details for evaluating the term $\mathcal{I}_1$ are provided in appendix \ref{details}. 
Here we list out the contribution of each of the 7 terms. 
\begin{eqnarray}
\mathcal{I}_1 &=&\frac{(\mathcal{E} + 4\mathcal{C}- 2 \mathcal{D})}{2}
\left(-\frac{149 \pi ^2 E^4}{9450}\right),
\\ \nonumber
\mathcal{I}_2 &=&  \frac{-(\mathcal{D}-2\mathcal{B})}{2}
\left(-\frac{29 \pi ^2 E^4}{5400}+\frac{19 \pi ^2 E^4}{37800}\right), \\
\mathcal{I}_3 &=& \frac{(\mathcal{D}-4\mathcal{C})}{2}
\left( -\frac{989 \pi ^2 E^4}{37800}\right),\nonumber\\
\mathcal{I}_4 &=&  \frac{\mathcal{C}}{2} \left(\frac{-11}{525} \pi ^2 E^4\right), \nonumber \\
\mathcal{I}_5 &=&
 \frac{ \mathcal{B}}{2}\left(\frac{127 E^4 \pi ^2}{75600}+\frac{91 E^4 \pi ^2}{10800}\right),\nonumber\\
\mathcal{I}_6 
&=& \frac{(\mathcal{B}-2\mathcal{A})}{2}\frac{29 E^4 \pi ^2}{10800},\nonumber\\
\mathcal{I}_7 &=& \frac{\mathcal{A}}{2} \frac{11 E^4 \pi ^2}{3600}. \nonumber
\end{eqnarray}

Summing all these terms together we get for the numerator in (\ref{ratiott}) 
\begin{eqnarray}
g^1_T(E) &=& -\frac{1}{180} (5 \mathcal{A}+7 \mathcal{B}-24 \mathcal{C}) \pi ^2 E^4.
\end{eqnarray}
Evaluating the denominator we obtain
\begin{eqnarray}
g^2_T(E) &=& -\frac{1}{180} (10 \mathcal{A}-2 \mathcal{B}-16 \mathcal{C}) \pi ^3 E^3.
\end{eqnarray}
Therefore the $(11)$ matrix element is given by 
\begin{eqnarray}\label{et11}
\hat E(T)_{11} &=& \frac{E}{\pi}\frac{ (5 \mathcal{A}+7 \mathcal{B}-24 \mathcal{C})}{10 \mathcal{A}-2 \mathcal{B}-16 \mathcal{C}}, \nonumber\\
&=& \frac{E}{4\pi} (1-\frac{t_4}{4}), 
\end{eqnarray}
where $t_4$  is defined by 
\begin{eqnarray}\label{t4def}
t_4 &\equiv& -\frac{4 (30 \mathcal{A}+90 \mathcal{B}-240 \mathcal{C})}{3 (10 \mathcal{A}-2 \mathcal{B}-16 \mathcal{C})}= -\frac{4 (n^T_f-n^T_s)}{n^T_f+n^T_s}.
\end{eqnarray}

Let us now examine the second diagonal element of the energy 
matrix $\hat E(T)$. For this we choose the polarisations to be 
\begin{eqnarray}
&&\epsilon^{xx}_1 = -\epsilon^{yy}_1 = 1, \qquad \epsilon^{--}_2 = 1,\nonumber\\
&&\epsilon^{xx}_3 = -\epsilon^{yy}_3= 1.\
\end{eqnarray}
The matrix element is given by 
\begin{eqnarray}
\hat E(T)_{22} 
&=& \frac{h_1(E)}{h_2(E)}
\end{eqnarray}
where
\begin{eqnarray}
h_1(E) &=& \int d^3x e^{iEt}\; \lim_{x_1^+ \rightarrow \infty}\frac{x_1^+ - x_1^-}{4}
\int_{-\infty}^{\infty} {dx_1^-}  \times  \\ 
&&  \qquad\qquad
\langle  \left(T_{xx}(x)-T_{yy}(x)\right) T_{--}(x_1)\left(T_{xx}(0)-T_{yy}(0)\right) \rangle,
\nonumber \\ \nonumber
h_2(E) &=& \int d^3x e^{iEt}\;
\langle \left(T_{xx}(x)-T_{yy}(x)\right)\left(T_{xx}(0)-T_{yy}(0)\right) \rangle.
\end{eqnarray}
Proceeding identically we obtain
\begin{eqnarray}
h_1(E) &=&\frac{1}{2} \frac{16}{45} (\mathcal{B}-2 \mathcal{C}) \pi ^2 E^4, \nonumber\\
h_2(E) &=& -\frac{4}{180} (10 \mathcal{A}-2\mathcal{B}-16 \mathcal{C}) \pi ^3 E^3. 
\end{eqnarray}
Evaluating the ratio we get
\begin{eqnarray}\label{et22}
\hat E(T)_{22}&=& \frac{E}{4\pi}\left(-\frac{16 (\mathcal{B}-2 \mathcal{C})}{5 \mathcal{A}-\mathcal{B}-8 \mathcal{C}}\right) ,\nonumber\\
&=&\frac{E}{4\pi}(1+\frac{t_4}{4}).
\end{eqnarray}
Note that if one  restricts the theories to be only parity even and require  that 
the energy is positive results. Then from (\ref{et11}) and (\ref{et22}) 
we obtain   the constraint
\begin{equation}
|t_4| \leq 4,
\end{equation}
which agrees with the results of \cite{Buchel:2009sk}.

\subsection{Parity odd contribution}

The off diagonal elements in the energy matrix receive contributions 
only from the parity odd terms in the three point function (\ref{ttt}). 
The parity odd term consists of 6 terms which are defined as 

\begin{eqnarray}\label{tttparityodd}
\langle T(x)T(x_1)T(0) \rangle_{\rm{odd}}
&=& p_T\frac{(P_1^2Q_1^2+5P_2^2P_3^2)S_1+(P_2^2Q_2^2+5P_3^2P_1^2)S_2+(P_3^2Q_3^2+5P_3^2P_1^2)S_3}{|x-x_1||x_1||-x|},\nonumber\\
&\equiv&p_T( I_1^p+I_2^p+I_3^p+I_4^p+I_5^p+I_6^p).
\end{eqnarray}

To obtain the $(12)$ element of the energy matrix we choose the 
polarisations to be given by 
\begin{eqnarray}
&&\epsilon_1^{xy} = \epsilon_1^{yx} =  \frac{1}{2}, 
\qquad \epsilon_3^{xx} = -\epsilon_3^{yy}=1,\nonumber\\
&&\epsilon_2^{--} = 1.
\end{eqnarray}
We  symmetrise the $(x,y)$ 
  component of the polarisation tensor $\epsilon_1^{xy}$
  \begin{eqnarray}
  \epsilon^{\mu\nu}_1 T_{\mu\nu} &=& \frac{1}{2}(T_{xy} + T_{yx}).
  \end{eqnarray}
  Then the required matrix element can be written as the ratio
\begin{eqnarray}
\hat E(T)_{12}  &=& \frac{f_1(E)}{f_2(E)},
\end{eqnarray}
where,
\begin{eqnarray}
f_1(E) &=& \int d^3x e^{iEt}\; \lim_{x_1^+ \rightarrow \infty}\frac{x_1^+ - x_1^-}{8}\int_{-\infty}^{\infty} {dx_1^-} (T_{xy} + T_{yx})(x) T_{--}(x_1)(T_{xx}(0)-T_{yy}(0)),\nonumber\\
&=& \frac{p_T}{2}(\mathcal{I}^p_1+\mathcal{I}^p_2+\mathcal{I}^p_3+\mathcal{I}^p_4+\mathcal{I}^p_5+\mathcal{I}^p_6),\nonumber \\
\end{eqnarray}
and 
\begin{eqnarray}
\mathcal{I}^p_n &=& \int d^3x e^{iEt}\; \lim_{x_1^+ \rightarrow \infty}\frac{x_1^+ - x_1^-}{4}\int_{-\infty}^{\infty} {dx_1^-} I^p_n.
\end{eqnarray}
To evaluate the denominator we use the two point function of the stress tensor 
(\ref{2pttt}) and 
it results in
\begin{eqnarray}\label{den12}
f_2(E) &=& -\frac{1}{90} (10 \mathcal{A}-2 \mathcal{B}-16 \mathcal{C}) \pi ^3 E^3.
\end{eqnarray}
As an example the contribution ${\cal I}^p_1$ to the numerator is evaluated in 
the appendix \ref{details}. 
Here we write down the result for each of the contributions
\begin{eqnarray}
\mathcal{I}^p_{1} &=&-  \frac{1}{256}\frac{8}{315} \pi ^3 E^4, \\ \nonumber
\mathcal{I}^p_{2} &=&\frac{1}{256} \left(\frac{1}{7} -\frac{4}{63} \right) \pi ^3 E ^4, \\ \nonumber
\mathcal{I}^{p}_{3} 
&=& \frac{1}{256} \left(\frac{4 }{105} - \frac{22}{315}   - \frac{4}{105} 
 + \frac{2 }{315}\right)\pi^3E^4, \\ \nonumber
\mathcal{I}^{p}_{4} 
&=& \frac{1}{256} \left( -\frac{4}{63}  -\frac{2}{21}\right) \pi ^3 E^4, \\ \nonumber
\mathcal{I}^{p}_{5} 
 &=& \frac{1}{256} \left( -\frac{1}{105}    -\frac{29}{315} \right) \pi ^3 E^4  , \\ \nonumber
\mathcal{I}^{p}_{6} 
 &=& -\frac{1}{256} \frac{25}{63}  \pi ^3  E ^4.
\end{eqnarray}
Summing all the contributions to the numerator we obtain
\begin{eqnarray}\label{num12}
f^1_T(E) &=& \frac{p_T}{2}(\mathcal{I}^{p}_{1}+\mathcal{I}^{p}_{2}+\mathcal{I}^{p}_{3}+\mathcal{I}^{p}_{4}+\mathcal{I}^{p}_{5}+\mathcal{I}^{p}_{6}),\nonumber\\
&=& -\frac{ p_T}{3\times 256}  \pi ^3  E^4.
\end{eqnarray}  
Using (\ref{den12}) and (\ref{num12}) we see that the $(12)$ element of the energy 
matrix is given by 
\begin{eqnarray}
\hat E(T)_{12}  &=& \frac{f^1_T(E)}{f^2_T(E)}, \nonumber\\
 &=& \frac{p_T}{256}\frac{15 E }{5\mathcal{A}-\mathcal{B}-8 \mathcal{C}}, \nonumber\\
 &=& \frac{E}{16 \pi} \alpha_T,
\end{eqnarray}  
where we define $\alpha_T$ by 
\begin{eqnarray}\label{alphaTdef}
\alpha_T &\equiv& \frac{p_T}{256}\frac{240 \pi }{ 5\mathcal{A}-\mathcal{B}-8 \mathcal{C}}=\frac{8 \pi ^4 p_T}{3 (n^T_f+n^T_s)}.
\end{eqnarray}

Finally let us examine the second off diagonal element of the 
energy matrix. We choose the polarisations to be given by
\begin{eqnarray}
& & \epsilon_3^{xy} = \epsilon_3^{yx} =\frac{1}{2}, 
\qquad  \epsilon_1^{xx} = -\epsilon_1^{yy}=1,\nonumber\\
& & \epsilon_2^{--} = 1.
\end{eqnarray}
Again this matrix element can be written as the ratio
\begin{eqnarray}\label{et21}
\hat E(T)_{21} &=& \frac{f'^1_T(E)}{f^2_T(E)},
\end{eqnarray}
where the numerator is given by 
\begin{eqnarray}
f'^1_T(E) &=& \int d^3x e^{iEt}\; \lim_{x_1^+ \rightarrow \infty}\frac{x_1^+ - x_1^-}{8}\int_{-\infty}^{\infty} {dx_1^-}  (T_{xx}(x)-T_{yy}(x))T_{--}(x_1) (T_{xy}(0)+ T_{yx}(0)). \nonumber\\
\end{eqnarray}
The denominator in (\ref{et21}) is the same as given in 
(\ref{den12}).
Evaluating the numerator using the same methods we have shown  that 
the numerator  is given by
\begin{eqnarray}
f'^1_T(E) &=& f^1_T(E)= -\frac{\alpha}{256}\frac{1}{3}  \pi ^3  E^4. \nonumber\\
\end{eqnarray}
This result coincides with the numerator of the $(12)$ element given in 
(\ref{num12}). 
This implies that 
\begin{eqnarray}
\hat E(T)_{12}  &=&  \hat E(T)_{21}.
\end{eqnarray}  
As mentioned earlier for the energy matrix corresponding to the
charge excitations, the fact that the off diagonal entries of the matrix
coincide serve a  simple check of our calculations.

 Using all the results we can write the energy matrix for 
 the excitations created by the stress tensor to be given by 
 \begin{eqnarray}
 \hat E(T) = \begin{pmatrix}
 \frac{E}{4\pi} (1-\frac{t_4}{4}) &&\frac{E}{16 \pi} \alpha_T \\
 \\
\frac{E}{16 \pi} \alpha_T&& \frac{E}{4\pi}(1+\frac{t_4}{4})\\
 \end{pmatrix},
\end{eqnarray} 
where,
\begin{eqnarray}\label{entrytmat}
t_4 &=& -\frac{4 (30 \mathcal{A}+90 \mathcal{B}-240 \mathcal{C})}{3 (10 \mathcal{A}-2 \mathcal{B}-16 \mathcal{C})} = -\frac{4 (n^T_f-n^T_s)}{n^T_f+n^T_s},\nonumber\\
\alpha_T &=& \frac{p_T}{256}\frac{240 \pi }{5 \mathcal{A}-\mathcal{B}-8 \mathcal{C}} = \frac{8 \pi ^4 p_T}{3 (n^T_f+n^T_s)}.
\end{eqnarray}
The condition that the energy observed at the calorimeter of the conformal 
collider is positive leads to the fact that the eigen values of the matrix $\hat E(T)$ are 
positive. The trace is positive, which implies that the determinant has to be positive. 
This leads to the constraint 
\begin{eqnarray}\label{causalityconstraints}
t_4^2 + \alpha_T^2 \leq 16.
 \end{eqnarray}
 Thus the theory dependent  parameters $t_4, \alpha_T$ of  the three point 
 function are constrained to lie on a disc of radius $4$ at the origin. 
 All three-dimensional conformal field theories which satisfy the 
 conformal collider bounds of \cite{Hofman:2008ar} lie in this disc. 
 The constraint (\ref{causalityconstraints})  was also obtained in 
 Appendix C of \cite{Maldacena:2011jn} by writing down the 
 energy observed at the conformal collider purely from symmetry arguments. 
 However the precise relation (\ref{entrytmat})  between $\alpha_T$ and the 
 coefficient $p_T$  of the parity odd term in the three point function 
 of the stress tensor was not given.

\section{Large $N$ Chern Simons theories}\label{largeNchernsimons}

$U(N)$ Chern Simons theory at level $\kappa$ coupled to either fermions or bosons
 in the fundamental 
representation \cite{Giombi:2011kc,Aharony:2012nh} are examples of conformal field theories which violate parity. In the large $N$ limit, these can be solved to all orders in the 't Hooft coupling 
\footnote{Note that $\kappa$ is the level defined using dimensional regularization, and differs from the level $k$ defined using Yang-Mills regularization by $\kappa=k+N$. $|\kappa|>N$ hence $|\lambda|\leq 1$.}
\begin{equation}
\lambda=\frac{N}{\kappa}.
\end{equation}
Lets write the three point functions of interest again
\begin{eqnarray}\label{3ptfn}
\langle jjT \rangle &=& n^j_s\langle jjT \rangle_{\textrm{free boson}} 
+n^j_{f}\langle jjT \rangle_{\textrm{free fermion}} + p_j \langle jjT \rangle_{\textrm{parity odd}},\\ 
\langle TTT \rangle &=& n^T_s\langle TTT \rangle_{\textrm{free boson}} 
+n^T_{f}\langle TTT \rangle_{\textrm{free fermion}} + p_T \langle TTT \rangle_{\textrm{parity odd}}.\nonumber  
\end{eqnarray}
where $\langle .. \rangle_{\textrm{free boson}}$, 
$\langle .. \rangle_{\textrm{free fermion}}$ denote the correlator of   a single 
real free  boson and a single real 
free fermion respectively.  
The theory dependent coefficients $n_s^{j, T}, n_f^{j, T}$ and $p_{j, T}$ 
in the three point functions (\ref{3ptfn})
are functions of the 't Hooft coupling which can be determined using 
softly broken higher spin symmetry \cite{Maldacena:2012sf} or direct calculation \cite{Giombi:2011kc,Aharony:2012nh,GurAri:2012is}. 
These coefficients for $U(N)$  Chern-Simons theory coupled
to fermions in the fundamental representation are given by
\begin{eqnarray}\label{cscoeffTTTim}
n^T_s(f) = n^j_s(f)  = 2N\frac{\sin \theta}{ \theta} \sin^2 \frac{\theta}{2}, &\qquad &
 n^T_f(f) = n^j_s(f) = 2N\frac{\sin \theta}{ \theta} \cos^2 \frac{\theta}{2},  \nonumber \\
 p_j(f)= \alpha' N \frac{\sin^2 \theta}{\theta} , &\qquad &
p_T(f) = \alpha N \frac{\sin^2 \theta}{\theta},
\end{eqnarray}
where the  t 'Hooft coupling is related to $\theta$ by 
\begin{equation}
\theta = \frac{\pi N}{\kappa}.
\end{equation}
Given the normalisation of the 
parity odd tensor structure in (\ref{jjt}) and 
(\ref{ttt}), the numerical coefficients $\alpha, \alpha'$ can be determined by 
a one loop computation either in the theory with fundamental fermions or
in theory with  fundamental bosons. 
In this section we  present the results of this perturbative calculation. 
This  was  done  earlier in
\cite{Giombi:2011kc}, we repeat the analysis to precisely determine the 
factors $\alpha, \alpha'$. 

Before we proceed, lets examine the coefficients 
for the parity even terms in (\ref{cscoeffTTTim}).
Note that we have chosen the normalisation of the stress tensor $T$
and the current $j$ in (\ref{3ptfn}) to agree with that given in 
\cite{Osborn:1993cr}.  This is evident from the 
tensor structures we have used for the parity even part in (\ref{jjt}) and (\ref{ttt}) and  
the normalisations of the  two point functions in (\ref{2ptjj}) and (\ref{2pttt}). 
Also observe that 
 taking the free limit $\theta\rightarrow 0$, 
we see that the three point function must reduce to that of $2N$ decoupled 
real fermions. This is clear from the limit 
$n_f^{j, T} \rightarrow 2N$ as $\theta \rightarrow 0$. 
To fix the factor in front the parity even bosonic contribution, we use the fact that we
can obtain the three point functions of the (critical) bosonic theory by the duality transformation
\cite{Aharony:2012nh,GurAri:2012is}
\begin{equation}
n_s^{j, T}(f)\rightarrow n_s^{j, T} (b) = 2N_b \frac{\sin \theta_b}{\theta_b} 
\cos^2 \frac{\theta_b}{2}.
\end{equation}
Now taking the limit $\theta_b\rightarrow 0$ we see 
that $n_s^{j, T}(b) \rightarrow  2N_b$.

\subsection{Parity odd three point functions}\label{podd3pt}

The parity odd coefficients $\alpha, \alpha'$ can be determined
by performing a one loop computation of $\langle jjT\rangle$ and 
$\langle TTT\rangle$ in the  $U(N)$ Chern-Simons theory 
coupled to fundamental fermions. 
 At one-loop these correlators are necessarily parity odd. While these calculations were previously done in \citep{Giombi:2011kc} \footnote{A one-loop calculation of  $\langle jjT \rangle$ in Chern-Simons coupled to fundamental bosons also appears  in \citep{Aharony:2011jz}.}, we explicitly redo the calculation 
here for completeness, and to ensure the precise numerical factors are correct
\footnote{The numerical factors $\alpha, \alpha'$ differ slightly from the one-loop calculations in \cite{Giombi:2011kc, Aharony:2011jz}.}.
 The calculation closely follows appendix G of \citep{Giombi:2011kc}. 

The action for the theory can be written in Euclidean space as 
\begin{equation}
S = \frac{i\kappa}{4 \pi}\int
\text{Tr} \left(A d A + \frac{2}{3} A^3\right)+ \int d^3x
 {\bar \psi}\gamma^\mu D_\mu  \psi,
\end{equation}
where
$$D_\mu \psi = \partial_\mu \psi -i A^a T^a \psi,$$ 
and the generators $T_a$ are normalized as: $\text{Tr } T_a^2=\frac{1}{2}$.
It is convenient to work in the gauge $A_3=0$. We have the following propagators:
\begin{eqnarray}
\langle A^a_i (x) A^b_j(0) \rangle & = & \frac{2\pi i}{\kappa} \epsilon_{ij} \text{sign}(x_3) \delta^2(\vec{x})\delta^{ab}, \\
\langle \psi^n(x) \bar{\psi}_m(0) \rangle & = & \frac{1}{4\pi} \frac{x_\mu \gamma^\mu}{|x|^3} \delta^n_m.
\end{eqnarray}
Here the gauge field $A_\mu=A_\mu^aT^a$, where $T^a$ are generators normalised by 
\begin{equation} 
\sum_{a} (T^a)_m^n (T^a)_p^q=\frac{1}{2} \delta_m^q \delta_p^n.
\end{equation}
Indices $i$ and $j$ can take on the values $1$ and $2$. It is also convenient to define Euclidean light cone coordinates as: $x_{\mp}=x^\pm=\frac{1}{\sqrt{2}}(x^1\pm i x^2)$.

We will calculate correlation functions with all free indices in the $x_-$ direction. 
We define the normalisations of $\tilde T_{--}$ and $j_-$ as
\begin{eqnarray}
j_- & = & \bar{\psi} \gamma_- \psi, \\
\tilde T_{--} & = & \bar{\psi} \gamma_- (\overrightarrow{D}_- - \overleftarrow{D}_-) \psi = \bar{\psi} \gamma_- \overleftrightarrow{D}_- \psi.
\end{eqnarray}
Note that here the normalisation of the stress tensor is twice
that of \cite{Osborn:1993cr} which is  what we have been 
using in the rest of the paper. 
\begin{equation} \label{stressnorm}
\tilde T = 2 T.
\end{equation}
We will incorporate this change in normalisation 
towards the end of our calculations. 
With these normalizations, we have, in free theory,
\begin{eqnarray}
 \langle J_-(x) J_-(0) \rangle & = & -N \frac{1}{4 \pi^2} \frac{x_-^2}{x^6} = -N \frac{1}{4 \pi^2}  \frac{P_3^2}{x^2},\\
  \langle \tilde T_{--}(x) \tilde T_{--}(0) \rangle & = & N \frac{3}{\pi^2} \frac{x_-^4}{x^{10}} =  N \frac{3}{\pi^2}\frac{ P_3^4}{x^2}.
\end{eqnarray}
These are also valid at one-loop, since the two-point functions must be parity even.
We will calculate all our correlation functions at the following points: 
\begin{eqnarray}\label{posi}
 x_1 & = & (x_1^+, x_1^-, x_1^3) = (0,0,0), \\
 x_2 & = & -(\delta^+, \delta^-,0), \\
 x_3 & = & (0,0,t),
\end{eqnarray}
and we will assume that 
\begin{equation}\label{sqlimit}
|\delta| \ll t,  \qquad \hbox{and} \quad  t>0.
\end{equation}

\begin{figure}[h]
\center
\includegraphics[width=4cm]{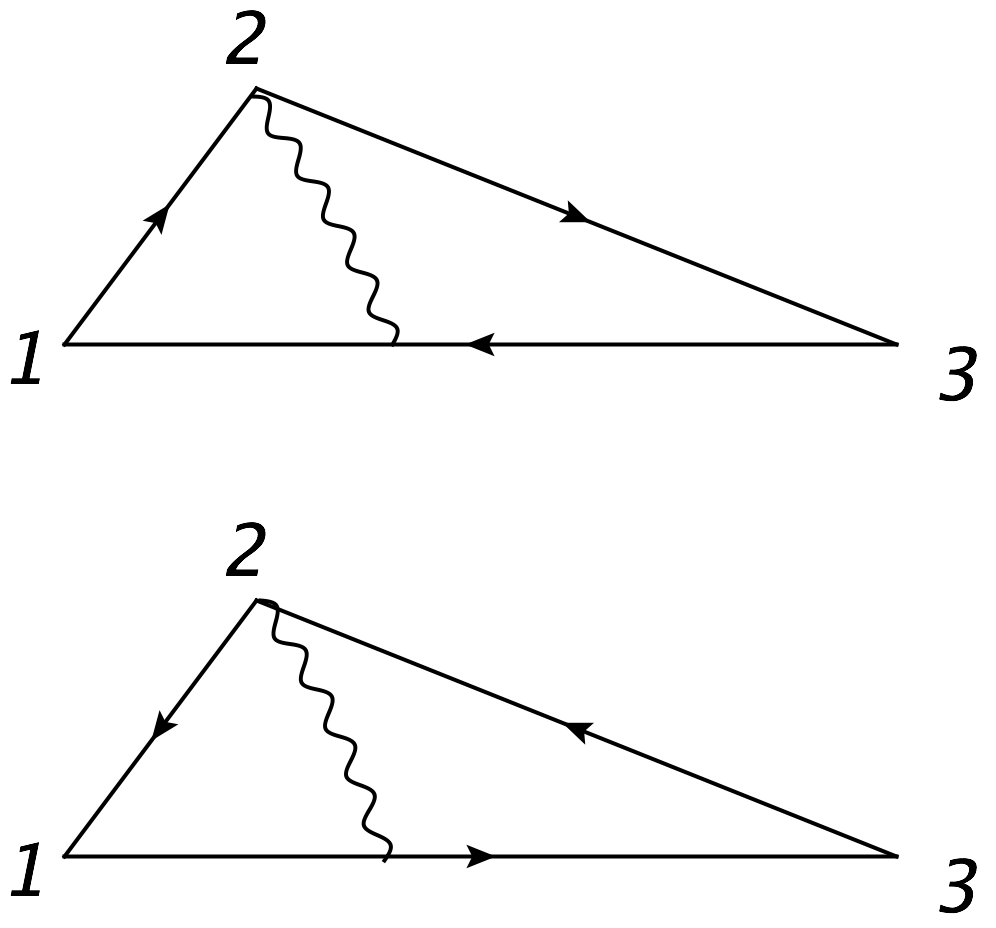}
\caption{Diagram IA and IB}\label{diagram1}
\end{figure}

\begin{figure}[h]
\center
\includegraphics[width=4cm]{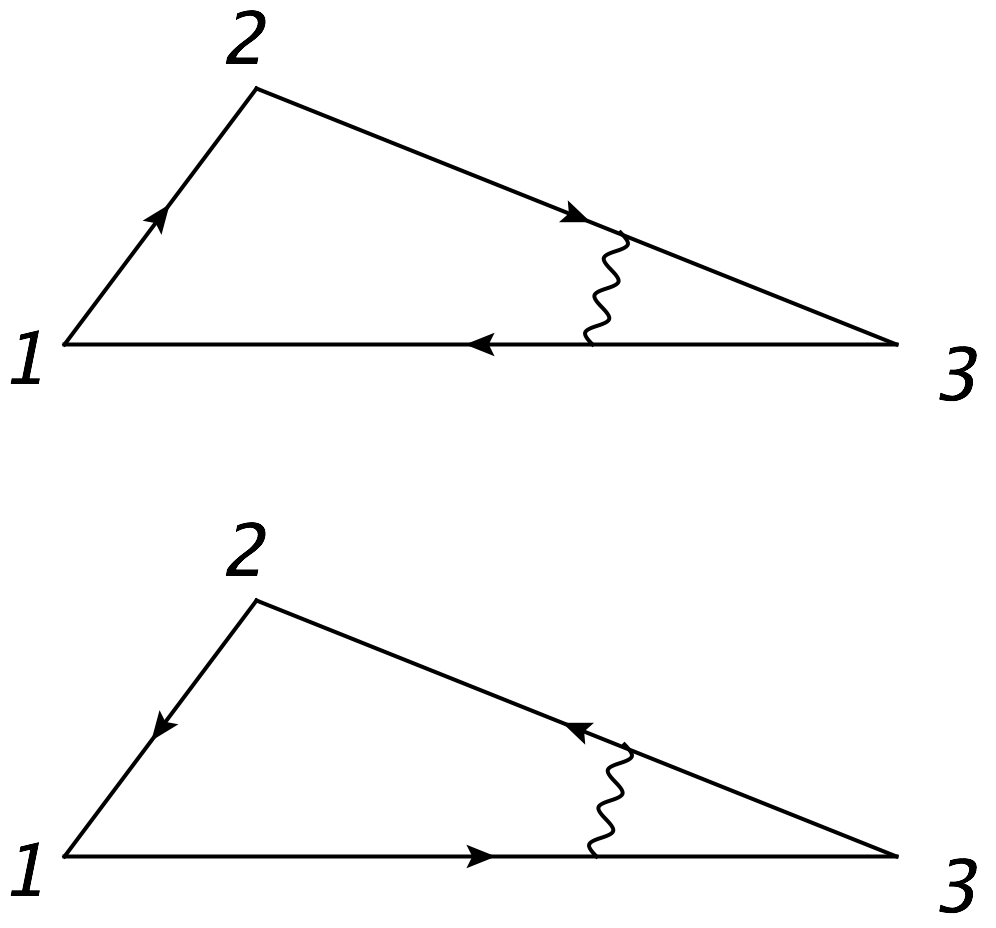}
\caption{Diagram IIA and IIB}\label{diagram2}
\end{figure}

\begin{figure}[h]
\center
\includegraphics[width=4cm]{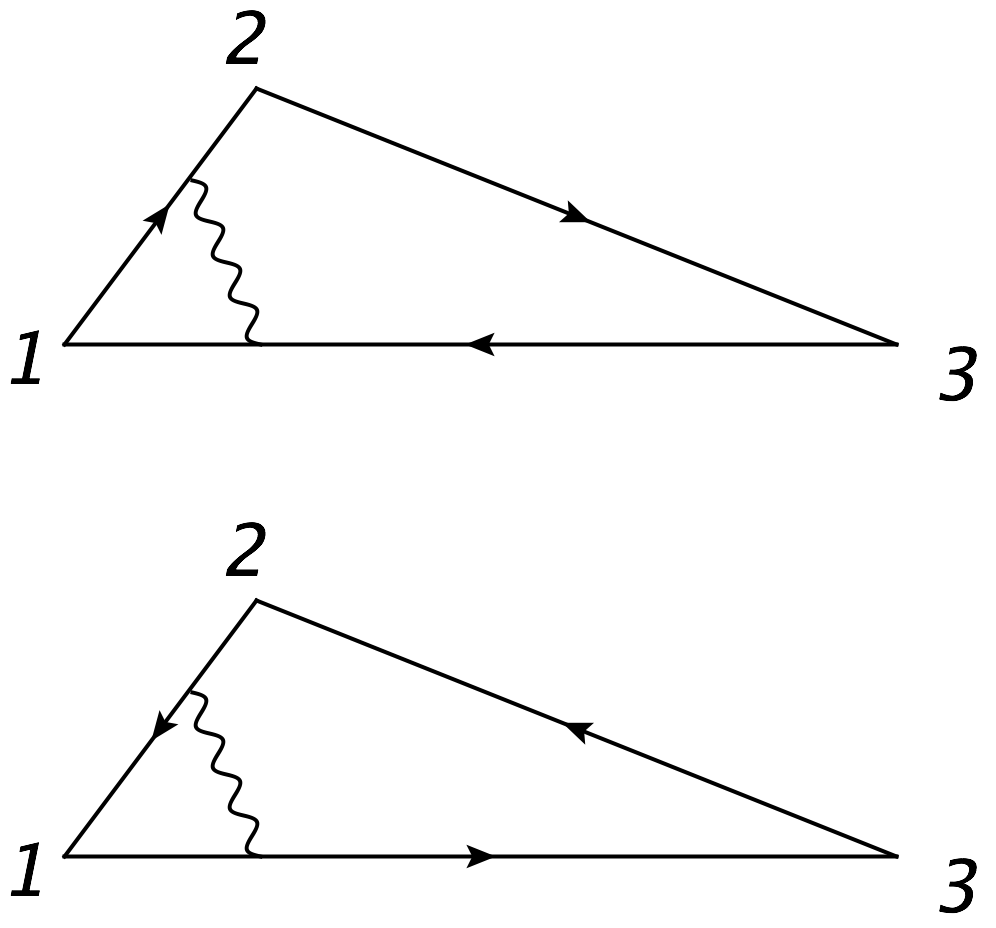}
\caption{Diagram IIIA and IIIB}\label{diagram3}
\end{figure}

With this particular choice of points,  temporal gauge and all free indices on operators in the same null direction, many diagrams vanish. The only diagrams which are 
non-vanishing are given in figures \ref{diagram1}, \ref{diagram2} and \ref{diagram3}.
After carrying out the calculation, we find that, the dominant contributions in our limit are from 
diagram IIIA and IIIB given in figure \ref{diagram3}. Other diagrams are 
sub-leading in our limit. (There are also some artefacts of temporal gauge 
produced by each diagram, which cancel amongst each other. 
See appendix G of \citep{Giombi:2011kc} for more details.) 
In the appendix \ref{oneloopcalculation} we present the calculations only of diagram III. 

\subsection*{Summary of the results}

Note that, using definitions\footnote{Note that the definitions of $P_i^2$, $Q_i^2$ and $S_i$ given in \cite{Giombi:2011kc} and \citep{Giombi:2011rz} differ by a factor of 4.} in \citep{Giombi:2011kc}, in the limit given by 
(\ref{posi}) and (\ref{sqlimit})  
\begin{eqnarray}\label{sqlimit1}
 P_3^2 & = &  \frac{(x_{12}^+)^2}{x_{12}^4} =  \frac{(\delta^+)^2}{\delta^4},\\
 P_1^2 & = &  \frac{(x_{23}^+)^2}{x_{23}^4} =  \frac{(\delta^+)^2}{(t^2+\delta^2)^2} \sim  \frac{(\delta^+)^2}{t^4}, \nonumber \\
 P_2^2 & = &  \frac{(x_{31}^+)^2}{x_{31}^4} = 0 , \nonumber \\
 Q_1 & = & - \frac{\delta^+}{\delta^2} , \qquad
 Q_2  =  0, \qquad
 Q_3  =  \frac{\delta^+}{t^2} , \nonumber  \\ 
 S_1  &=&  i \frac{(\delta^+)^2}{4\delta t^3}, \qquad
 S_2  =  0 ,   \qquad
 S_3  = - i \frac{(\delta^+)^2 t}{4\delta^3 t^2}.   \nonumber
\end{eqnarray}

Using the limits (\ref{sqlimit1}), the odd tensor structure for $\langle jj\tilde T\rangle$ reduces to 
\begin{eqnarray}\label{tenlimit}
 \left. \frac{1}{|x_{12}||x_{23}||x_{31}|} \left(Q_3^2 S_3 + 2 P_1^2S_2 +2 P_2^2 S_1\right)
 \right|_{|\delta/t \rightarrow| 0} = -i \frac{\delta_-^4}{4\delta^4 t^7}.
\end{eqnarray}
In appendix \ref{oneloopcalculation} we have 
evaluated  the diagrams in figure \ref{diagram3} and we find 
\begin{eqnarray}\label{normalisationjjt}
\lim_{ |{\delta}/{t}| \rightarrow 0}
\langle jj\tilde T \rangle_{\rm{odd}} &=& N \lambda \frac{1}{2 \pi^3} \frac{\delta_-^4}{\delta^4 t^7}.
\end{eqnarray}

Therefore using (\ref{tenlimit}),   the parity odd part of the three point function is given by
\begin{eqnarray}
\langle jj \tilde T\rangle_{{\rm odd}}
= N \lambda \frac{2i}{\pi^3} \ \frac{\left(Q_3^2 S_3 + 2 P_1^2S_2 +2 P_2^2 S_1\right)}{|x_{12}||x_{23}||x_{31}|} .
\end{eqnarray}
Now going back to Minkowski signature and taking into account of 
the normalisation (\ref{stressnorm}) we obtain
\begin{equation} 
\langle jj  T\rangle_{{\rm odd}}
= N \lambda \frac{1}{\pi^3} \ \frac{\left(Q_3^2 S_3 + 2 P_1^2S_2 +2 P_2^2 S_1\right)}{|x_{12}||x_{23}||x_{31}|}. 
\end{equation}
Finally let us compare  the above equation with the normalisation for $p_{j}$
 written in (\ref{cscoeffTTTim}). For this we need to take the $\theta\rightarrow 0$ limit. 
 We see that
 \begin{equation} \label{defalp}
 \alpha' = \frac{1}{\pi^4}.
 \end{equation}

Let us perform the same analysis  for the parity odd part of the 
three point function of the stress tensor.  
In the limit (\ref{sqlimit1}) the parity odd tensor structure reduces to 
\begin{eqnarray}\label{tenlimitt}
  && \left. \frac{1}{|x_{12}||x_{23}||x_{31}|}\left[ (P_1^2 Q_1^2 +5 P_2^2P_3^2)S_1 + (P_2^2 Q_2^2 +5 P_3^2P_1^2)S_2 + (P_3^2 Q_3^2 +5 P_1^2P_2^2)S_3\right] 
  \right|_{|\delta/t \rightarrow| 0}\nonumber  \\
  &  & \qquad\qquad\qquad \qquad\qquad = - i \frac{\delta_-^6}{4\delta^8 t^7} .
\end{eqnarray}
Evaluating  the corresponding diagram in figure \ref{diagram3} in appendix 
we obtain 
\begin{equation}\label{normalisationttt}
    \langle \tilde T\tilde T\tilde T \rangle = N \lambda \frac{6}{\pi^3}\frac{(\delta^+)^6}{\delta^8 t^7} .
  \end{equation}
Therefore using (\ref{tenlimitt}) we can write the parity odd part of the
three point function of the stress tensor as
\begin{eqnarray}
& &\langle \tilde T\tilde T\tilde T \rangle_{{\rm odd}} = N \lambda \frac{24i}{\pi^3} \times 
\\ \nonumber
& & 
\frac{1}{|x_{12}||x_{23}||x_{31}|}\left[ (P_1^2 Q_1^2 +5 P_2^2P_3^2)S_1 + (P_2^2 Q_2^2 +5 P_3^2P_1^2)S_2 + (P_3^2 Q_3^2 +5 P_1^2P_2^2)S_3\right] .
\end{eqnarray}
In minkowski signature and using the normalisation of the stress tensor in 
(\ref{stressnorm}) we obtain 
\begin{eqnarray}
& &\langle  T T T \rangle_{\rm odd} = N \lambda \frac{3}{\pi^3} \times 
\\ \nonumber
& & 
\frac{1}{|x_{12}||x_{23}||x_{31}|}\left[ (P_1^2 Q_1^2 +5 P_2^2P_3^2)S_1 + (P_2^2 Q_2^2 +5 P_3^2P_1^2)S_2 + (P_3^2 Q_3^2 +5 P_1^2P_2^2)S_3\right] .
\end{eqnarray}
Finally we fix the normalisation of the coefficient $p_T$ in (\ref{cscoeffTTTim}) by 
taking the $\theta\rightarrow 0 $ limit. We obtain
\begin{equation}\label{defal}
\alpha = \frac{3}{\pi^4}.
\end{equation}

\subsection{Saturation of the conformal collider bounds}

Given values for the coefficients $n_s^{j, T}, n_f^{j, T}$ and $p_{j, T}$ 
given in (\ref{cscoeffTTTim}) with $\alpha, \alpha' $ as evaluated in (\ref{defal}) and
(\ref{defalp}) we can evaluate the  entries of the energy matrix 
corresponding to the charge and stress tensor excitations. 
Before we present the final result, for reference we list 
here the intermediate parameters ${\cal A}, {\cal B}, {\cal C}$ which appear
in the three point function of the stress tensor.
We use the relations  \cite{Osborn:1993cr}  
\begin{eqnarray}
\mathcal{A} &=& \frac{27 n_s^T}{512 \pi ^3}, \qquad \mathcal{B} = -\frac{9 (4 n_f^T + 3n_s^T )}{512 \pi ^3}, \qquad \mathcal{C} = -\frac{9 (8 n_f^T + n_s^T)}{2048 \pi ^3},
\end{eqnarray}
to obtain
\begin{eqnarray}
\mathcal{A}&=& \frac{27 N \sin \frac{\theta }{2}^2 \sin \theta }{256 \pi ^3 \theta }, \nonumber\\
\mathcal{B}&=&  -\frac{9 N \sin \theta  (\cos \theta +7)}{512 \pi ^3 \theta }, \nonumber\\ 
\mathcal{C}&=& -\frac{9 N \sin \theta  (7 \cos \theta +9)}{2048 \pi ^3 \theta }. 
\end{eqnarray}  
Similarly the  intermediate parameters $c, e$, that occur in the 3 point function of the 
stress tensor with 2 insertions of the current is related to the 
coefficients  in (\ref{cscoeffTTTim}) by
\begin{eqnarray}
c &=& \frac{3 (2 n^j_f + n^j_s)}{256 \pi ^3}, \qquad
e = \frac{3 n^j_s}{256 \pi ^3} .
\end{eqnarray}
This leads to 
\begin{eqnarray}
c &=& \frac{3 N (3+\cos \theta ) \sin \theta }{256 \pi ^3 \theta },\\ \nonumber
e &=& \frac{3 N \sin  \frac{\theta }{2} ^2 \sin \theta }{128 \pi ^3 \theta }.
\end{eqnarray}

Now using (\ref{cscoeffTTTim}) with $\alpha, \alpha' $ as evaluated in (\ref{defal}) and
(\ref{defalp})  and the 
relations (\ref{entryjmat})  and (\ref{entrytmat}) we obtain
\begin{eqnarray}
a_2 = -2 \cos\theta,  \qquad \alpha_j = 2 \sin\theta, \qquad
t_4 = -4 \cos\theta, \qquad  \alpha_T = 4 \sin\theta.
\end{eqnarray}
Thus the Chern-Simons theories with a single fundamental boson or fermion saturate the 
conformal collider bounds and they lie on the circles
\begin{equation}
a_2^2 + \alpha_j^2 = 4, \qquad t_4^2 + \alpha_T^2 = 16.
\end{equation}
The location of the theory on the circle is 
given by $\theta$ the t 'Hooft coupling.

We note that, in the large $N$ limit, Chern-Simons theories coupled to fundamental bosons or fermions are very similar to free theories, because they also contain an infinite tower of higher spin operators, whose scaling dimensions ($\Delta_s=s+1$) saturate the unitarity bound ($\Delta_s \geq s+1$), hence it is perhaps not surprising that they saturate the conformal collider bounds as well. 
The fact that these theories saturate the bounds  implies that 
there exist a polarisation of the energy matrix for which the eigen value vanishes. 
It will be interesting to show this directly from the conservation of higher spin 
currents in these theories. 
Such a proof will demonstrate that conservation of higher spin currents 
ensures that the conformal collider bounds will be saturated.
Large $N$ Chern-Simons theory coupled to more general types of fundamental matter, such as theories with both a fundamental boson and fermion \cite{Jain:2013gza}, as well as supersymmetric theories, or ABJ in the limit $M\ll N$ \cite{Chang:2012kt}
also contain a tower of higher-spin operators at the unitarity bound. 
Let us consider the case of large $N$ Chern-Simons theory  with both 
fundamental boson and fermion, in such a theory there exists two sets of 
$U(1)$ as well as spin $2$ currents. 
Therefore the polarisation at which  the energy vanishes are different
for each of currents. The combined currents will therefore not 
have a  polarisation at which the energy vanishes
\footnote{We thank Ofer Aharony for raising these points.}.
%
%

On the other hand, we might expect that strongly-interacting large $N$ Chern-Simons theories coupled to matter in other representations, such as adjoint or bi-fundamental matter  \cite{Aharony:2008ug, Banerjee:2013nca, Guru:2014ad}, where the higher-spin operators are not at the unitarity bound, must lie inside the disc, away from the boundary. Correlation functions in pure Einstein gravity, which are of course, parity-preserving, lie at the centre of the disk. However, perhaps that one could find a counter example to this using parity-violating gravity theories with an axion \cite{Bayntun:2010nx, Goldstein:2010aw}.

\section{Conclusions}

We have obtained constraints on the three-point functions $\langle jjT\rangle, 
\langle TTT\rangle$  that apply to all
(both parity-even and parity-odd) conformal field theories in $d=3$. 
These constraints were obtained by imposing the condition
that energy observed at the conformal collider be positive. 
The constraints we obtain imply that 
the space of all allowed $\langle jjT\rangle, 
\langle TTT\rangle$ correlation functions in  conformal field theories in $d=3$ lie
 on a two-dimensional disc. These constraints are particularly relevant for Chern-Simons theories with matter, and we explicitly showed that the $\langle jjT\rangle, 
\langle TTT\rangle$ correlation functions of large $N$, $U(N)$ Chern-Simons theories
 with a single fundamental fermion or a fundamental boson
 lie on the bounding circles of these disc.

 In this paper we have restricted our analysis to excitations created by 
 the $U(1)$ current and the stress tensor. However we expect similar 
 results for excitations  carrying arbitrary spin. It will be interesting to
 generalise the observations of this paper to these correlation functions.

 Recent evaluation of parity violating three point functions in certain
 higher spin theories Vasiliev theories in $AdS_4$  show that these theories 
 are dual to 
  large $N$ matter Chern-Simons theories \cite{Sezgin:2017jgm,Didenko:2017lsn}. 
  Our results then imply that these Vasiliev like theories also 
  saturate the conformal collider bounds and lie on the 
  circles that bound the disc in the parameter space of the 
  the three point functions. 
  It will be interesting to show this directly from a shock wave type 
   analysis as was done in $AdS_5$ by \cite{Hofman:2016awc}. 
  
  Another direction to explore is to obtain a proof of the conformal collider
   bounds for parity odd theories in $d=3$
  from unitarity   and causality 
  of the CFT along the lines of 
  \cite{Hartman:2015lfa,Hartman:2016lgu,Hofman:2016awc,Hartman:2016dxc,Dymarsky:2017xzb}. For this 
  we need to understand the structure of parity odd conformal blocks
  in $d=3$ which in itself is an useful enterprise.
 
\acknowledgments
We thank Ofer Aharony for reading the manuscript and useful correspondence. 
S.D.C would like to thank  Alexander Zhiboedov, D. M. Hofman for discussions. S.D.C is grateful to ICTP, Trieste for organising Spring School on Superstring Theory and Related Topics, 2017 and also acknowledges the hospitality provided by IFT Madrid, Lorentz Institute for theoretical physics, Leiden and IOP, Amsterdam . S.P was supported in part by 
an INSPIRE award of the Department of Science and Technology, India. S.P. would like to thank 
the International Centre for Theoretical Sciences (ICTS), Bengaluru and the Department of 
Theoretical Physics, Tata Institute of Fundamental Research, Mumbai for hospitality.

\appendix 

\section{Details for evaluation of  the energy matrix $\hat E(T)$}
\label{details}

In this appendix we will illustrate the procedure involved in evaluating the 
contributions to the energy matrix $\hat E(T)$ . 

We first detail the steps for 
 the term ${\cal I}_1$ that occurs in 
the parity even element $(11)$ of the energy matrix $\hat E(T)$ defined in (\ref{ttterms}). 
All the other terms are evaluated similarly. 
We need to first take the $x_1^+\rightarrow\infty$ limit, then the 
integral over the null time $x_1^-$ and finally  perform the integral over 
the $3$ remaining spatial directions.  The steps are  outlined in the following 
equations. 
\begin{eqnarray}
\mathcal{I}_1 &=& \int d^3x e^{iEt}\; \lim_{x_1^+ \rightarrow \infty}\frac{x_1^+ - x_1^-}{2}\int_{-\infty}^{\infty} \frac{dx_1^-}{2} I_1, \nonumber\\
&=& (\mathcal{E} + 4\mathcal{C}- 2 \mathcal{D})\int d^3x e^{iEt}\; \lim_{x_1^+ \rightarrow \infty}\frac{x_1^+ - x_1^-}{2}\int_{-\infty}^{\infty} \frac{dx_1^-}{2} \frac{\mathcal{I}^T_{xy, \mu' \nu'}(x)\mathcal{I}^T_{--, \sigma' \rho'}(x_1)}{x^6 x_1^6} \nonumber\\
&&\left(\frac{Z^{\mu'}Z^{\nu'}}{Z^2}-\frac{1}{3} \eta^{\mu' \nu'}\right)\left(\frac{Z_{x}Z_{y}}{Z^2}\right)\left(\frac{Z^{\sigma'}Z^{\rho'}}{Z^2}-\frac{1}{3} \eta^{ \sigma' \rho'} \right)\frac{1}{Z^{\frac{3}{2}}},\nonumber\\
&=& (\mathcal{E} + 4\mathcal{C}- 2 \mathcal{D})\int d^3x e^{iEt}\; \lim_{x_1^+ \rightarrow \infty}\frac{x_1^+ - x_1^-}{2}\int_{-\infty}^{\infty} \frac{dx_1^-}{2} \frac{(\mathcal{I}^T_{x+, \mu' \nu'}(x)-\mathcal{I}^T_{x-, \mu' \nu'}(x))\mathcal{I}^T_{--, \sigma' \rho'}(x_1)}{x^6 x_1^6} \nonumber\\
&&\left(\frac{Z^{\mu'}Z^{\nu'}}{Z^2}-\frac{1}{3} \eta^{\mu' \nu'}\right)\left(\frac{Z_{x}(Z_{+}-Z_-)}{Z^2}\right)\left(\frac{Z^{\sigma'}Z^{\rho'}}{Z^2}-\frac{1}{3} \eta^{ \sigma' \rho'} \right)\frac{1}{Z^{\frac{3}{2}}},\nonumber\\
&=& (\mathcal{E} + 4\mathcal{C}- 2 \mathcal{D})\int d^3x e^{iEt}\; \nonumber\\
&&\int_{-\infty}^{\infty} \frac{dx_1^-}{2} \frac{x^2 (x^-)^2 \sqrt{\frac{-x^-+x_1^-}{\left(x^2-x^- x^+\right) x_1^-}} \left(x^2-(x^-)^2+(x^--x^+) x_1^-\right) \left(x^2+x^+ x_1^--x^- (x^++x_1^-)\right)}{32 \left(x^2-(x^--2i\epsilon) (x^+-2i\epsilon)\right)^4 (x^--x_1^-+i\epsilon)^4 (x_1^--i\epsilon)^3},\nonumber\\
&=& (\mathcal{E} + 4\mathcal{C}- 2 \mathcal{D})\int d^3x e^{iEt}\; \nonumber\\
&&\int_{-\infty}^{\infty} \frac{dx_1^-}{2} \frac{(x^2(x^-)^2)}{32\left(x^2-x^- x^+\right)^\frac{9}{2}}\left( \frac{x^4-x^2 (x^-)^2-x^2 x^- x^++(x^-)^3 x^+
}{(x_1^--i\epsilon)^\frac{7}{2}(x_1^--x^-+i\epsilon)^\frac{7}{2}} + \frac{-(x^-)^2+2 x^- x^+-(x^+)^2}{(x_1^--i\epsilon)^\frac{3}{2}(x_1^--x^-+i\epsilon)^\frac{7}{2}}\right.\nonumber\\
&&\left. + \frac{(x^-)^3+x^- x^+-2 (x^-)^2 x^+}{(x_1^--i\epsilon)^\frac{5}{2}(x_1^--x^-+i\epsilon)^\frac{7}{2}} \right), \nonumber\\
&=&\frac{(\mathcal{E} + 4\mathcal{C}- 2 \mathcal{D})}{2} \nonumber\\
&&\int d^3x e^{iEt} \left(\frac{2 x^2 \left(16 x^4-16 x^2 x^- (x^-+x^+)+(x^-)^2 \left(5 (x^-)^2+6 x^- x^++5 (x^+)^2\right)\right)}{15 (x^--2i\epsilon)^4 \left(x^2-(x^--2i\epsilon)(x^+-2i\epsilon)\right)^{9/2}}\right).\nonumber\\
\end{eqnarray}
In the  last step
we have used integrals \eqref{F3}, \eqref{F6} ,\eqref{F1}. 
To perform the final integrations we
 introduce the light cone coordinates $x^\pm = t \pm y$. The integrations are  done by 
 first integrating over the direction orthogonal to the light cone directions ($x$) first
  and then the light cone directions. The steps are indicated below
\begin{eqnarray}
\mathcal{I}_1 &=&\frac{(\mathcal{E} + 4\mathcal{C}- 2 \mathcal{D})}{2} \frac{1}{2}\int_{-\infty}^\infty e^{\frac{iEx^+}{2}}\int_{-\infty}^\infty dx^- e^{\frac{iEx^-}{2}}\nonumber\\
&& \int_{-\infty}^\infty dx \left(\frac{2 x^2 \left(16 x^4-16 x^2 x^- (x^-+x^+)+(x^-)^2 \left(5 (x^-)^2+6 x^- x^++5 (x^+)^2\right)\right)}{15 (x^--2i\epsilon)^4 \left(x^2-(x^--2i\epsilon)(x^+-2i\epsilon)\right)^{9/2}}\right),\nonumber\\
&=& -\frac{(\mathcal{E} + 4\mathcal{C}- 2 \mathcal{D})}{2} \frac{1}{2}\int_{-\infty}^\infty e^{\frac{iEx^+}{2}}\int_{-\infty}^\infty dx^- e^{\frac{iEx^-}{2}} \frac{1504 \sqrt{-x^- x^+} \sqrt{-\frac{1}{x^- x^+}}}{1575 \left(x^--2i\epsilon\right)^5 (x^+-2i\epsilon)}\nonumber\\
&&+\frac{64 \sqrt{-x^- x^+} \sqrt{-\frac{1}{x^- x^+}}}{175 \left(x^--2i\epsilon\right)^4 \left(x^+-2i\epsilon\right)^2}+\frac{32 \sqrt{-x^- x^+} \sqrt{-\frac{1}{x^- x^+}}}{315 \left(x^--2i\epsilon\right)^3 \left(x^+-2i\epsilon\right)^3}, \nonumber\\
&=& \frac{(\mathcal{E} + 4\mathcal{C}- 2 \mathcal{D})}{2}(-\frac{149 \pi ^2 E^4}{9450}).
\end{eqnarray}
We have used \eqref{F4}, for integrating over the light cone directions.

As an illustration of the steps involved in obtaining the parity odd element
$(12)$ of the energy matrix $\hat E(T)$ we examine the term
corresponding to $I^p_1$ in (\ref{tttparityodd}). 
\begin{eqnarray}
I^p_1 &=& \frac{P_1^2Q_1^2S_1}{|x-x_1||x_1||-x|}|_{\epsilon_1^{xy} =\epsilon_1^{yx}= \epsilon_2^{--} = \epsilon_3^{xx} =1} + \frac{P_1^2Q_1^2S_1^2}{|x-x_1||x_1||-x|}|_{\epsilon_1^{xy} = \epsilon_1^{yx}  \epsilon_2^{--} =-\epsilon_3^{yy}=1}, \nonumber\\
&=& \frac{2}{256}\frac{-32 Q_{1x}Q_{1y}I_{-x}(x_1)}{x_1^6(x-x_1)^2x^2} \left( \epsilon^{\gamma'\delta}_{\phantom{\mu\nu}-} x_{1\gamma'} (x-x_1)_{\delta} x_{1x} - \frac{\epsilon^{\gamma'}_{\phantom{\mu}-x}}{2}\left( |x-x_1|^2x_{1\gamma'} + |x_1|^2 (x-x_1)_{\gamma'} \right) \right) \nonumber\\
&&+ \frac{2}{256}\frac{32 Q_{1x}Q_{1y}I_{-y}(x_1)}{x_1^6(x-x_1)^2x^2} \left( \epsilon^{\gamma'\delta}_{\phantom{\mu\nu}-} x_{1\gamma'} (x-x_1)_{\delta} x_{1y} - \frac{\epsilon^{\gamma'}_{\phantom{\mu}-y}}{2}\left( |x-x_1|^2x_{1\gamma'} + |x_1|^2 (x-x_1)_{\gamma'} \right) \right), \nonumber\\
&\equiv& I^p_{1,1} + I^p_{1,2}
\end{eqnarray}
The term $I^p_{1,1}$ does not contribute to the final result since in the 
limit $x_1^+ \rightarrow \infty$, it is sub-leading.
\begin{eqnarray}
& &\mathcal{I}^p_{1,1} = \int d^3x e^{iEt}\; \lim_{x_1^+ \rightarrow \infty}\frac{x_1^+ - x_1^-}{2}\int_{-\infty}^{\infty} \frac{dx_1^-}{2} I^p_{1,1},\nonumber\\
&=&  \int d^3x e^{iEt}\;\int_{-\infty}^{\infty} \frac{dx_1^-}{2}\lim_{x_1^+ \rightarrow \infty}\frac{x_1^+ - x_1^-}{2} \nonumber\\
&& \frac{2}{256}\frac{-32 Q_{1x}Q_{1y}I_{-x}(x_1)}{x_1^6(x-x_1)^2x^2} \left( \epsilon^{\gamma'\delta}_{\phantom{\mu\nu}-} x_{1\gamma'} (x-x_1)_{\delta} x_{1x} - \frac{\epsilon^{\gamma'}_{\phantom{\mu}-x}}{2}\left( |x-x_1|^2x_{1\gamma'} + |x_1|^2 (x-x_1)_{\gamma'} \right) \right), \nonumber\\
&=& \int d^3x e^{iEt}\;\int_{-\infty}^{\infty} \frac{dx_1^-}{2} O(\frac{1}{x_1^+}), \nonumber\\
&=& O(\frac{1}{x^+}).\nonumber\\
\end{eqnarray}
The term $I^p_{1, 2}$ is handled as before in the following steps.
\begin{eqnarray}
&&\mathcal{I}^p_{1,2} = \int d^3x e^{iEt}\; \lim_{x_1^+ \rightarrow \infty}\frac{x_1^+ - x_1^-}{2}\int_{-\infty}^{\infty} \frac{dx_1^-}{2} I^p_{1,2},\nonumber\\
&=&\frac{1}{256}\int d^3x e^{iEt}\;\int_{-\infty}^{\infty} {dx_1^-}\frac{2x^2 \left(x^2-(x^-)^2+x^- x_1^--x^+ x_1^-\right)}{2 \left(-x^2+(x^--2i\epsilon)(x^+-2i\epsilon)\right)^3 (x^--x_1^-+i\epsilon)^2 (x_1^--i\epsilon)^4}, \nonumber\\
&=&\frac{1}{256}\int d^3x e^{iEt}\;\int_{-\infty}^{\infty} {dx_1^-} \frac{-2}{2 \left(x^2-(x^--2i\epsilon)(x^+-2i\epsilon)\right)^3}(\frac{x^2 \left(x^2-(x^-)^2\right)}{(x_1^--x^- + i\epsilon)^2(x_1^--i\epsilon)^4} \nonumber\\
&& + \frac{x^2 (x^--x^+)}{(x_1^--x^- + i\epsilon)^2(x_1^--i\epsilon)^3}),\nonumber\\
&=& \frac{1}{256} \int d^3x e^{iEt} \frac{2i \pi  x^2 \left(4 x^2-x^- (x^-+3 x^+)\right)}{(x^--2i\epsilon)^5 \left(-x^2+(x^--2i\epsilon)(x^+-2i\epsilon)\right)^3}, \nonumber\\
&=& \frac{1}{256} \frac{1}{2} \int_{-\infty}^{\infty} dx^+ e^{\frac{iEx^+}{2}} \int_{-\infty}^{\infty} dx^- e^{\frac{iEx^-}{2}} \frac{2\pi ^2}{8 (x^--2i\epsilon)^{9/2} (x^+-2i\epsilon)^{3/2}}+\frac{30 \pi ^2}{8 (x^--2i\epsilon)^{11/2} \sqrt{x^+-2i\epsilon}}, \nonumber\\
&=&\frac{1}{256} (-\frac{8}{315} \pi ^3 E^4).
\end{eqnarray}

We have used  the integrals \eqref{F5}, \eqref{F2} and \eqref{F4}
\begin{eqnarray}
\mathcal{I}^p_{1} &=& -\frac{1}{32} \frac{1}{315} \pi ^3 E^4
\end{eqnarray}

\section{Parity odd three-point functions at one loop }\label{oneloopcalculation}

In this appendix we detail the steps involved in evaluating 
diagrams $IIIA, IIIB$ in figure \ref{diagram3}  for both the  $\langle jj\tilde T\rangle$ and the 
$\langle \tilde T \tilde T \tilde T\rangle$ correlator.

\subsection{Perturbative calculation of $\langle jj \tilde T \rangle$}

The integral corresponding to diagram IIIA is 
\begin{equation}
\begin{split}
    &
(\text{III}_A) = \int d^3y_1 d^3y_2 \langle A_i(y_1) A_j(y_2) \rangle {\rm Tr} \left[\gamma^i\langle \psi(y_1)\bar\psi(x_1)\rangle
{\gamma_-} \langle \psi(x_1)\bar\psi(y_2)\rangle \gamma^j \langle \psi(y_2)\bar\psi(x_2)\rangle \right.
\\
&\left.~~~~\times\gamma_-\langle \psi(x_2)\bar\psi(x_3)\rangle \gamma_-(\overleftrightarrow\partial_{x^-_3}) \langle \psi(x_3)\bar\psi(y_1)\rangle \right].
\end{split}
\end{equation}
The overall minus sign comes from the fermion loop.
Processing this we have:
\begin{equation}
\begin{split}
&\int d^3 y_1 d^3 y_2 \text{ sign}(y_{12}^0) \delta^2(\vec y_{12}) \epsilon_{ij} {\rm Tr}\left[ \gamma^i \slashed{y}_1 \gamma^+ (-\slashed{y}_2)\gamma^j (\slashed{y}_2+\slashed{\delta}) \gamma^+ (-\slashed{x}-\slashed{\delta}) \gamma^+ (\slashed{x}-\slashed{y}_1) \right]
\\
&~~~~\times   {1\over |y_1|^3 |y_2|^3 |\delta+y_2|^3}   \left( {1\over |x+\delta|^3} \overleftrightarrow\partial_{x^-}  {1\over |x-y _1|^3}\right) \left(\frac{1}{4\pi}\right)^5 \frac{2\pi i}{k} \left( \frac{N^2}{2} \right)
\\
& = \frac{N^2}{64 \pi^4 k} \delta^+ \int d^2z dt_1 dt_2 \text{ sign}(t_{12}) [(z^+)^2 t_2^2+z^+(z+\delta)^+t_1(t-t_1)]
\\
&~~~~\times   {1\over (z^2+t_1^2)^{3\over 2} (z^2+t_2^2)^{3\over 2} [(z+\delta)^2+t_2^2]^{3\over 2}}   \left\{  {1\over (t^2+\delta^2)^{3\over 2}} (\overleftarrow \partial_{\delta^-}+ \overrightarrow\partial_{z^-}) {1\over [(t-t_1)^2+z^2]^{3\over 2}}  \right\}.
\end{split}
\end{equation}
The trace over gamma matrices is carried out explicitly in the section below. The factor of $N^2/2$ is a group theory factor associated, with the factor of $1/2$ arising from the normalisation of the generators $T^a$. 

Diagram IIIB is the same as diagram IIIA with $x_2 \leftrightarrow x_3$: 
\begin{equation}
\begin{split}
    &
(\text{III}_B) = \int d^3y_1 d^3y_2 \langle A_i(y_1) A_j(y_2) \rangle {\rm Tr} \left[\gamma^i\langle \psi(y_1)\bar\psi(x_1)\rangle
{\gamma_-} \langle \psi(x_1)\bar\psi(y_2)\rangle \gamma^j \langle \psi(y_2)\bar\psi(x_3)\rangle \right.
\\
&\left.~~~~\times\gamma_- (\overleftrightarrow\partial_{x^-_3})\langle \psi(x_3)\bar\psi(x_2)\rangle \gamma_- \langle \psi(x_2)\bar\psi(y_1)\rangle \right].
\end{split}
\end{equation}
Processing this diagram, we see that, after evaluating the trace, it is identical to Diagram IIIA (with $t_1$ and $t_2$ interchanged):
\begin{equation}
\begin{split}
& = -\frac{N^2}{64 \pi^4 k} \delta^+ \int d^2z dt_1 dt_2 \text{ sign}(t_{12}) [(z^+)^2 t_1^2+z^+(z+\delta)^+t_2(t-t_2)]
\\
&~~~~\times   {1\over (z^2+t_1^2)^{3\over 2} (z^2+t_2^2)^{3\over 2} (t_1^2+(\delta+z)^2)^{3\over 2}}   \left\{ {1\over [(t_2-t)^2+z^2]^{3\over 2}} (\overleftarrow \partial_{z^-}+ \overrightarrow\partial_{\delta^-}) {1\over (t^2+\delta^2)^{3\over 2}} \right\}.
\end{split}
\end{equation}

Now summing both the diagrams we obtain
\begin{eqnarray}
 &&\text{III}_A+\text{III}_B  =  \frac{N^2}{32 \pi^4 k} \delta^+ \int d^2z dt_1 dt_2 \text{ sign}(t_{12}) [(z^+)^2 t_2^2+z^+(z+\delta)^+t_1(t-t_1)]
\\
&~& ~~~\times   {1\over (z^2+t_1^2)^{3\over 2} (z^2+t_2^2)^{3\over 2} [(z+\delta)^2+t_2^2]^{3\over 2}}   \left\{  {1\over (t^2+\delta^2)^{3\over 2}} (\overleftarrow \partial_{\delta^-}+ \overrightarrow\partial_{z^-}) {1\over [(t-t_1)^2+z^2]^{3\over 2}}  \right\}.\nonumber\\
\end{eqnarray}
This  includes all group theory factors and both diagrams. 

The leading contribution to these integrals comes from the regions D1: $z \sim O(\delta)$, $t_1, t_2 \sim O(\delta)$, and D2:$z \sim O(\delta)$, and $t-t_1 \sim O(\delta)$ $t_2 \sim O(\delta)$.
Evaluating the integral in the region $D2$ gives rise to an un-physical artefact of temporal gauge, which cancels with another diagram. See \citep{Giombi:2011kc} for details. For the final answer, therefore, we only require the contribution from region D1, which is:
\begin{equation}
\begin{split}
(\text{III}_A+\text{III}_B) \Big|_{D1} = -\frac{3N^2}{32\pi^4 k}  \frac{\delta^+}{t^7}  \int d^2z dt_1 dt_2 { \text{ sign}(t_{12})  z^+(z^+ + \delta^+)^2 t_1
\over (z^2+t_1^2)^{3\over 2} (z^2+t_2^2)^{3\over 2} [(z+\delta)^2+t_2^2]^{3\over 2}}.   
\end{split}
\end{equation}
Using
\begin{equation}
\int dt_1 \text{ sign}(t_1-t_2) \frac{t_1}{(z^2+t_1^2)^{3/2}} =  \frac{2}{(z^2+t_2^2)^{1/2}} ,
\end{equation}
we are left with an integral over $d^2z$ and $t_2$. Defining $z_3=t_2$, we find it can be written as:
\begin{equation}
\begin{split}
I \Big|_{D1} = -\frac{3N^2}{16\pi^4 k}  \frac{\delta^+}{t^7}  \int d^3z {   z^+(z^+ + \delta^+)^2 
\over  z^4 (z+\delta)^3} . 
\end{split}
\end{equation}
Evaluating this remaining integral over $z$ by Feynman parameters, we have:
\begin{equation}
\begin{split}
 \int d^3z {   z^+(z^+ + \delta^+)^2 
\over  z^4 (z+\delta)^3} = -\frac{8\pi}{3} \frac{\delta^+}{\delta^4} .  
\end{split}
\end{equation}

Therefore we obtain 
\begin{equation}
   \lim_{|\delta/t|\rightarrow 0} \langle jjT \rangle = \frac{N^2}{2 \pi^3 k} \frac{(\delta^+)^4}{\delta^4 t^7}. 
\end{equation}

\subsection{Perturbative calculation of $\langle \tilde T \tilde T \tilde T \rangle$}

Again, we focus on Diagram III (corrections to the $x_1$ vertex). As before
 there are two permutations that contribute:
\begin{equation}
\begin{split}
& \text{III}_A= \int d^3y_1 d^3y_2 \langle A_i(y_1) A_j(y_2) \rangle {\rm Tr} \left[\gamma^j\langle \psi(y_1)\bar\psi(x_1)\rangle
\gamma_- (\overleftrightarrow\partial_{x_1^-}) \langle \psi(x_1)\bar\psi(y_2)\rangle \gamma^i \langle \psi(y_2)\bar\psi(x_2)\rangle \right.
\\
&\left.~~~~\times\gamma_- (\overleftrightarrow\partial_{x_2^-}) \langle \psi(x_2)\bar\psi(x_3)\rangle \gamma_- (\overleftrightarrow\partial_{x_3^-}) \langle \psi(x_3)\bar\psi(y_1)\rangle \right].
\end{split}
\end{equation}
and
\begin{equation}
\begin{split}
& \text{III}_B= \int d^3y_1 d^3y_2 \langle A_i(y_1) A_j(y_2) \rangle {\rm Tr} \left[\gamma^j\langle \psi(y_1)\bar\psi(x_1)\rangle
\gamma_- (\overleftrightarrow\partial_{x_1^-}) \langle \psi(x_1)\bar\psi(y_2)\rangle \gamma^i \langle \psi(y_2)\bar\psi(x_3)\rangle \right.
\\
&\left.~~~~\times \gamma_- (\overleftrightarrow\partial_{x_3^-}) \langle \psi(x_3)\bar\psi(x_2)\rangle \gamma_- (\overleftrightarrow\partial_{x_2^-}) \langle \psi(x_2)\bar\psi(y_1)\rangle \right].
\end{split}
\end{equation}
As before, we see that they turn out to be identical. 

Diagram $IIIA$ can be written as:
\begin{equation}
\begin{split}
& C \int d^3 y_1 d^3 y_2 \text{ sign}(t_{12}) \delta^2(\vec y_{12}) \epsilon_{ij} {\rm Tr}\left[ \gamma^i \slashed{y}_1 \gamma^+ (-\slashed{y}_2)\gamma^j (\slashed{y}_2+\slashed{\delta}) \gamma^+ (-\slashed{x}-\slashed{\delta}) \gamma^+ (\slashed{x}-\slashed{y}_1) \right] 
\\
&~~~\times \left( {1\over |y_1|^3}(\overleftarrow{\partial}_{y_1^-}-\overrightarrow{\partial}_{y_2^-}) {1\over |y_2|^3} \right) \left[{1\over |\delta+y_2|^3}(\overleftarrow{\partial}_{\delta^-}-\overrightarrow{\partial}_{\delta^-}) {1\over |x+\delta|^3} (-\overleftarrow\partial_{x^-}+\overrightarrow\partial_{x^-}) {1\over |x-y _1|^3}\right]
\\
& = -16iC \delta^+ \int d^2z dt_1 dt_2 \text{ sign}(t_{12}) [(z^+)^2 t_2^2+z^+(z+\delta)^+t_1(t-t_1)]
\left[ {1\over (t_1^2+z^2)^{3\over 2}}\overleftrightarrow\partial_{z^-} {1\over (t_2^2+z^2)^{3\over 2}} \right]
\\
&~~~~\times  \left\{ {1\over [t_2^2+(\delta'+z)^2]^{3\over 2}} ( \overleftarrow\partial_{z^-} - \overrightarrow\partial_{\delta^-}) {1\over (t^2+\delta^2)^{3\over 2}}(-\overleftarrow\partial_{\delta^-}-\overrightarrow\partial_{z^-}) {1\over [(t-t_1)^2+z^2]^{3\over 2}} \right\}_{\delta'=\delta},
\end{split}
\end{equation}
where $C=\left(\frac{1}{4\pi}\right)^5 \frac{2\pi i}{k} \left( \frac{N^2}{2} \right)$.
This receives contributions from the two regions D1 and D2. Evaluating the region 
D2 gives rise to an un-physical artefact of temporal gauge which cancels with another diagram. 
The final answer for the correlation function in our limit thus only comes from region D1.
\begin{equation}
\begin{split}
& = -16iC {\delta^+t}{} \int d^2z dt_1 dt_2 \text{ sign}(t_{12}) [z^+(z+\delta)^+t_1]
\left[ {1\over (t_1^2+z^2)^{3\over 2}}\overleftrightarrow\partial_{z^-} {1\over (t_2^2+z^2)^{3\over 2}} \right]
\\
&~~~~\times  \left\{ {1\over [t_2^2+(\delta'+z)^2]^{3\over 2}} ( \overleftarrow\partial_{z^-}) {1\over (t^2+\delta^2)^{3\over 2}}(-\overleftarrow\partial_{\delta^-}-\overrightarrow\partial_{z^-}) {1\over [(t-t_1)^2+z^2]^{3\over 2}} \right\}, \\
& = -16iC \frac{\delta^+}{t^7} \int d^2z dt_1 dt_2 \text{ sign}(t_{12}) [z^+(z+\delta)^+t_1]
\left[ {1\over (t_1^2+z^2)^{3\over 2}}\overleftrightarrow\partial_{z^-} {1\over (t_2^2+z^2)^{3\over 2}} \right]
\\
&~~~~\times  \left\{ {-9(\delta^+ + z^+)^2  \over [t_2^2+(\delta+z)^2]^{5\over 2}}  \right\},
\\
& = -16iC \frac{\delta^+}{t^7} \int d^2z dt_1 dt_2 \text{ sign}(t_{12}) [z^+(z+\delta)^+t_1]
\left[ {-3z^+ (t_1^2-t_2^2)\over (t_1^2+z^2)^{5\over 2} (t_2^2+z^2)^{5\over 2}} \right]
\left\{ {-9(\delta^+ + z^+)^2  \over [t_2^2+(\delta+z)^2]^{5\over 2}}  \right\}.
\end{split}
\end{equation}
Now we use the integral
\begin{equation}
\int dt_1 \text{ sign}(t_{1}-t_2) t_1(t_1^2-t_2^2) (t_1^2+z^2)^{-5/2} = \frac{4}{3} \frac{1}{(t^2+z^2)^{1/2}},
\end{equation}
and let $z_3=t$, to obtain
\begin{equation}
\begin{split}
    &\text{III}_A = -16iC \frac{\delta^+}{t^7} \int d^3z 
 {36(\delta^+ + z^+)^3 (z^+)^2 \over z^6(\delta+z)^5}, \\
 & = 12\cdot 256\pi i C \frac{(\delta^+)^6}{\delta^8 t^7}.
\end{split}
\end{equation}
which was evaluated using Feynman parameters.

Diagram IIIB is
\begin{equation}
\begin{split}
& C \int d^3 y_1 d^3 y_2 \text{ sign}(t_{12}) \delta^2(\vec y_{12}) \epsilon_{ij} {\rm Tr}\left[ \gamma^i \slashed{y}_1 \gamma^+ (-\slashed{y}_2)\gamma^j (\slashed{y}_2-\slashed{x}) \gamma^+ (\slashed{x}+\slashed{\delta}) \gamma^+ (-\slashed{\delta}-\slashed{y}_1) \right] 
\\
&~~~\times \left( {1\over |y_1|^3}(\overleftarrow{\partial}_{y_1^-}-\overrightarrow{\partial}_{y_2^-}) {1\over |y_2|^3} \right) \left[{1\over |y_2-x|^3}(-\overleftarrow{\partial}_{x^-}+\overrightarrow{\partial}_{x^-}) {1\over |x+\delta|^3} (\overleftarrow\partial_{\delta^-}-\overrightarrow\partial_{\delta^-}) {1\over |\delta+y_1|^3}\right]
\\
& = -16iC \delta^+ \int d^2z dt_1 dt_2 \text{ sign}(t_{12}) [(z^+)^2 t_1^2+z^+(z+\delta)^+t_2(t-t_2)]
\left[ -{1\over (t_2^2+z^2)^{3\over 2}}\overleftrightarrow\partial_{z^-} {1\over (t_1^2+z^2)^{3\over 2}} \right]
\\
&~~~~\times  \left\{ {1\over [t_1^2+(\delta'+z)^2]^{3\over 2}} ( \overleftarrow\partial_{z^-} - \overrightarrow\partial_{\delta^-}) {1\over (t^1+\delta^2)^{3\over 2}}(-\overleftarrow\partial_{\delta^-}-\overrightarrow\partial_{z^-}) {1\over [(t-t_2)^2+z^2]^{3\over 2}} \right\}_{\delta'=\delta}.
\end{split}
\end{equation}
Using $\text{ sign}(t_{12})=
-\text{ sign}(t_{12})$ we see that this is equal to diagram IIIA (with $t_1$ and $t_2$ interchanged).

To summarize, the result is then:
\begin{equation}
\lim_{|\delta/t| \rightarrow 0} 
\langle \tilde T \tilde T\tilde T \rangle = \text{III}_A+\text{III}_B= 24\cdot 256\pi i C \frac{(\delta^+)^6}{\delta^8 t^7}={6 N^2 \over \pi^3 k} \frac{(\delta^+)^6}{\delta^8 t^7}.
\end{equation}

\subsection*{Traces over gamma matrices}
To evaluate the traces over gamma matrices, we use $\epsilon_{+-}=i$,
$(\gamma^+)^2=0$ and relations such as
\begin{equation}
\rm{Tr } (\gamma^{\alpha_1}\gamma^{\alpha_2}\gamma^{\alpha_3}\gamma^{\alpha_4}
) = 2 \delta^{\alpha_1 \alpha_2}\delta^{\alpha_3 \alpha_4} +2
\delta^{\alpha_1 \alpha_4}\delta^{\alpha_2 \alpha_3}-2
\delta^{\alpha_1 \alpha_3}\delta^{\alpha_2 \alpha_4},
\end{equation}
%
from which we find:
\begin{eqnarray}
 && i\text{Tr } \left( \gamma^+ \gamma^\mu \gamma^+ \gamma^\nu \gamma^- \gamma^\rho \gamma^+ \gamma^\sigma \gamma^+ \gamma^\eta \right), \\
 &=& 8 i \eta^{+\eta}\eta^{+\mu} \eta^{+\sigma} \text{Tr } \left( \gamma^+ \gamma^\nu \gamma^- \gamma^\rho \right), \\
& = & 16i \eta^{+\eta}\eta^{+\mu} \eta^{+\sigma} \left( \eta^{+\nu} \eta^{-\rho} + \eta^{+\rho}\eta^{-\nu}-\eta^{\nu \rho} \right).
\end{eqnarray}
and
\begin{eqnarray}
 && -i\text{Tr } \left( \gamma^- \gamma^\mu \gamma^+ \gamma^\nu \gamma^+ \gamma^\rho \gamma^+ \gamma^\sigma \gamma^+ \gamma^\eta \right), \\
 &=& -8 i \eta^{+\sigma}\eta^{+\nu} \eta^{+\rho} \text{Tr } \left( \gamma^- \gamma^\mu \gamma^+ \gamma^\eta \right), \\
& = & -16i \eta^{+\sigma}\eta^{+\nu} \eta^{+\rho} \left( \eta^{-\mu} \eta^{+\eta} + \eta^{-\eta}\eta^{+\mu}-\eta^{\mu \eta} \right).
\end{eqnarray}
These imply that 
\begin{eqnarray}
&&\epsilon_{ij} {\rm Tr}\left[ \gamma^i \slashed{y}_1 \gamma^+ (-\slashed{y}_2)\gamma^j (\slashed{y}_2+\slashed{\delta}) \gamma^+ (-\slashed{x}-\slashed{\delta}) \gamma^+ (\slashed{x}-\slashed{y}_1) \right] =  \\ \nonumber
&&\qquad\qquad\qquad\qquad16i \delta^+\left( (z^+)^2 t_2^2+z^+ (z^+ + \delta^+)  t_1(t-t_1) \right).
\end{eqnarray}
where, from the delta function in the propagator we can set $\vec{y}_1=\vec{y}_2=\vec{z}$, $x^3=t$, $y_1^3=t_1$ and $y_2^3=t_2$.

\section{Table of integrals}\label{A}

In this appendix we list out the various integrals used to evaluate the 
Energy matrix.  We also outline the steps in performing some of them.

\begin{eqnarray}\label{F1}
\int_{-\infty}^{\infty} dz \frac{1}{(z-i\epsilon)^d (z-x+i\epsilon)^{d+1}} 
&=&\int_{-\infty}^\infty dz \int_0^\infty ds \frac{(z -i\epsilon)s^{d}e^{\left(-s(z-i\epsilon)(z-x+i\epsilon)\right)}}{\Gamma[d +1]},\nonumber\\
&=&  \int_0^\infty ds \frac{e^{-\frac{1}{4} s (2 \epsilon+i x)^2} \sqrt{\pi } (-2 i \epsilon+x)}{2 \sqrt{s}} \frac{s^d}{\Gamma[d+1]},  \nonumber\\
&=& \frac{4^d \sqrt{\pi } \left((2 \epsilon+i x)^2\right)^{-\frac{1}{2}-d} (-2 i \epsilon +x) \Gamma\left[\frac{1}{2}+d\right]}{\Gamma[1+d]}. \nonumber\\ 
\end{eqnarray}

\begin{eqnarray}\label{F2}
\int_{-\infty}^{\infty} dz \frac{1}{(z-i\epsilon)^{d+1} (z-x+i\epsilon)^{d}} &=& \int_{- \infty}^\infty dz \int_0^\infty ds \frac{(z-x+i\epsilon)s^{d}e^{\left(-s(z-i\epsilon)(z-x+i\epsilon)\right)}}{\Gamma[d +1]},\nonumber\\
 &=& \frac{4^d \sqrt{\pi } (2 i \epsilon -x) \left((2 \epsilon +i x)^2\right)^{-\frac{1}{2}-d} \Gamma\left[\frac{1}{2}+d\right]}{\Gamma[1+d]}.
\end{eqnarray}

\begin{eqnarray}\label{F3} 
\int_{-\infty}^{\infty} dz \frac{1}{(z-i\epsilon)^{d} (z-x+i\epsilon)^{d}} &=& \frac{\sqrt{\pi } (-1)^{\frac{1}{2}-d} \left(2\right)^{-1+2d} \Gamma\left[-\frac{1}{2}+d\right]}{(x-2i\epsilon)^{2d}\Gamma[d]}.
\end{eqnarray} 
               
\begin{eqnarray}\label{F4}
\int_{-\infty}^{\infty} dz \frac{e^{\frac{iEz}{2}}}{(z -i\epsilon)^d} \nonumber
&=& \int_{0}^{\infty} dz \frac{e^{\frac{iEz}{2}}}{(z -i\epsilon)^d}  + \int_{-\infty}^{0} dz \frac{e^{\frac{iEz}{2}}}{(z -i\epsilon)^d} ,\nonumber\\
&=& \int_{0}^{\infty} dz \frac{e^{\frac{iEz}{2}}}{(z -i\epsilon)^d} + e^{i\pi d} \int_{0}^{\infty} dz \frac{e^{\frac{-iEz}{2}}}{(z +i\epsilon)^d}, \nonumber\\
&=& \int_{0}^{\infty} dz \int_0^\infty ds \frac{s^{d-1}}{\Gamma[d]} e^{-s(z-i\epsilon)+\frac{iEz}{2}} \nonumber\\
&&+ e^{i\pi d} \int_{0}^{\infty} dz \int_0^\infty ds \frac{s^{d-1}}{\Gamma[d]} e^{-s(z +i\epsilon)-\frac{iEz}{2}}, \nonumber\\
 &=& \frac{1}{\Gamma[d]} \int_0^\infty ds s^{d-1} e^{is\epsilon} \frac{1}{s-i\epsilon} + e^{i\pi d}\frac{1}{\Gamma[d]} \int_0^\infty ds s^{d-1} e^{-is\epsilon} \frac{1}{-s+i\epsilon},\nonumber\\
 &=& \frac{1}{\Gamma[d]} \int_{-\infty}^\infty ds s^{d-1} e^{is\epsilon} \frac{1}{s-i\epsilon}, \nonumber\\
 &=& \frac{2\pi i}{\Gamma(d)}\left(\frac{iE}{2}\right)^{d-1}.
\end{eqnarray}

\begin{eqnarray}\label{F5}
\int_{-\infty}^{\infty} dz \frac{1}{(z-i\epsilon)^{d+2} (z-x+i\epsilon)^{d}} &=& -\frac{2^{1+2 d} d \sqrt{\pi } \left((2 \epsilon +i x)^2\right)^{-\frac{1}{2}-d} \Gamma\left[\frac{1}{2}+d\right]}{\Gamma[2+d]}.
\end{eqnarray}

\begin{eqnarray}\label{F6}
\int_{-\infty}^{\infty} dz \frac{1}{(z-i\epsilon)^{d} (z-x+i\epsilon)^{d+2}} &=& -\frac{2^{1+2 d} d \sqrt{\pi } \left((2 \epsilon +i x)^2\right)^{-\frac{1}{2}-d} \Gamma\left[\frac{1}{2}+d\right]}{\Gamma[2+d]}.
\end{eqnarray}

\begin{eqnarray}\label{F7}
\int_{-\infty}^{\infty} dz \frac{1}{(z-i\epsilon)^{d} (z-x+i\epsilon)^{d+3}} &=& \frac{(2\pi i) (-1)^{-d-3} \Gamma[2d+2]}{\Gamma[d]\Gamma[d+3] (x^- - 2i\epsilon)^{2d+2}}.
\end{eqnarray}

\begin{eqnarray}\label{F8}
\int_{-\infty}^{\infty} dz \frac{1}{(z-i\epsilon)^{d+3} (z-x+i\epsilon)^{d}} &=& \frac{(2\pi i) (-1)^{-d} \Gamma[2d+2]}{\Gamma[d]\Gamma[d+3] (x^- - 2i\epsilon)^{2d+2}}.
\end{eqnarray}

\begin{eqnarray}\label{F9}
\int_{-\infty}^{\infty} dz \frac{1}{(z-i\epsilon)^{d+4} (z-x+i\epsilon)^{d}} &=& \frac{2^{3+2 d} d (1+d) \sqrt{\pi } \left((2 \epsilon +i x)^2\right)^{-\frac{3}{2}-d} \Gamma\left[\frac{3}{2}+d\right]}{\Gamma[4+d]}. \nonumber\\
\end{eqnarray}

\begin{eqnarray}\label{F10}
\int_{-\infty}^{\infty} dz \frac{1}{(z-i\epsilon)^{d} (z-x+i\epsilon)^{d+4}} &=& \frac{2^{3+2 d} d (1+d) \sqrt{\pi } \left((2 \epsilon +i x)^2\right)^{-\frac{3}{2}-d} \Gamma\left[\frac{3}{2}+d\right]}{\Gamma[4+d]}. \nonumber\\
\end{eqnarray}

\providecommand{\href}[2]{#2}\begingroup\raggedright\endgroup

\end{document}